\documentclass[useAMS,usenatbib]{mn2e}
\usepackage{epstopdf,amsmath,amsfonts,amssymb,graphicx,subfigure,enumitem,epstopdf}

\usepackage{natbib}
\bibpunct{(}{)}{;}{a}{,}{,}

\usepackage{graphicx}
\usepackage{helvet}
\usepackage{natbib}
\usepackage{setspace}
\usepackage{amssymb}
\usepackage{longtable}
\title[Variable stars]
{Photometric observations of NGC 281: Detection of 228 variable stars}

\author[Sneh Lata et al.]
       {Sneh Lata$^1$\thanks{E-mail: sneh@aries.res.in}, A. K. Pandey$^1$, J. C. Pandey$^1$, Neelam Panwar$^1$, Paulomi Paul$^1$ \\
       $^1$Aryabhatta Research Institute of Observational Sciences, Manora Peak, Nainital 263002, Uttarakhand, India \\}

\date{Accepted ---------.
      Received ---------;
      }

\pagerange{\pageref{firstpage}--\pageref{lastpage}}

\def\LaTeX{L\kern-.36em\raise.3ex\hbox{a}\kern-.15em
    T\kern-.1667em\lower.7ex\hbox{E}\kern-.125emX}

\begin{document}

\label{firstpage}

\maketitle

\label{firstpage}
\begin{abstract}

We identify 228 periodic variables in the region of young open cluster NGC 281 using time series photometry carried 
out from 1 m class ARIES telescopes, Nainital.
The cluster membership of these identified variables is determined on the basis colour-magnitude, two colour diagrams and kinematic data.
 Eighty one variable stars are found to be members of the cluster NGC 281. Of 81 variables,  
30 and 51 are probable main-sequence and pre-main-sequence members, respectively. 
Present study classifies main-sequence variable stars into different variability types according to their periods of variability,  
shape of light curves 
and location in the Hertzsprung-Russell diagram. These identified main-sequence variables could be $\beta$ Cep, $\delta$ Scuti, slowly pulsating B type 
and new class variables. Among 51 pre-main-sequence variables, majority of them are weak line T Tauri stars. 
The remaining 147 variables could belong to the field population. The variability characteristics of the field population indicate that these variables could be RR Lyrae, $\delta$ Scuti and binaries type variables.
\end{abstract}

\begin {keywords} 
Open cluster:  NGC 281  --
colour--magnitude diagram: Variables: pre-main-sequence stars; T Tauri stars 
\end {keywords}

\section{Introduction}
This work presents time series observations of young open cluster NGC 281 as a part of our ongoing project entitled “Search for pre-main-sequence (PMS) variability in Young
Open Clusters (YOCs)”.
YOCs contain significant number of PMS stars with circumstellar disk and are
unique laboratories to study the evolution of disks of PMS stars. 
 In general, PMS stars are young low-mass 
sources which remain deeply embedded the parent molecular cloud.
PMS stars show variability associated to a number of different physical processes. 
Most of the physical processes such as dark spots and flaring, etc are associated with magnetic and rotational activities (Herbst
 et al. 1994).
As a result, PMS
variability can be expressed over a wide range of timescales, from intra day to months.
These PMS stars accrete matter from their surrounded accretion disk and show prominent 
features such large IR excess and H${\alpha}$ emission line.
In addition to PMS stars, YOCs also contain several O, B and Be type stars which  are also 
found to be pulsating or rotating variables with timescales that range from minutes to years. 
Among these variables, there are about 20\% B-type stars which show Be phenomenon i.e. B star with emission line. 
Be type variables are very rapidly rotating main-sequence (MS) stars surrounded by a circumstellar envelope of gas. 
These stars show variability which is related to the presence of a circumstellar disk of variable size and 
structure (Arcos et al. 2018).  

\begin{figure*}
\includegraphics[width=17cm]{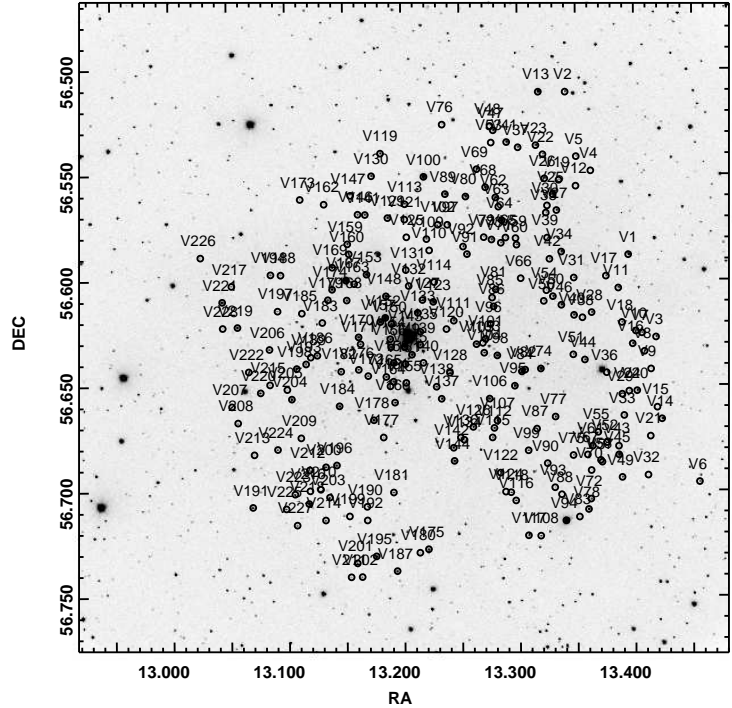}
\caption{The observed image in $V$ band which contains open cluster NGC 281. The stars detected as variables are encircled.  
 }
\end{figure*}
NGC 281 is a basically HII region known as Sh2-184 and located at a distance of 2.8 kpc (Elmegreen \& Lada 1978; Sato et al. 2008; Sharma et al. 2012).
The MS of this cluster is well defined and contains significant number of O and B spectral type stars.
The reddening $E(B-V)$ towards the cluster is found to be low $\sim$0.32 mag.
Previous studies suggest that star formation is still continued in this cluster, hence the cluster provides an excellent laboratory to search PMS for variable stars. A detailed multiwavelength study of the region containing NGC 281 was presented by Sharma et al. (2012). They found that the majority of the identified PMS stars have age $\sim$1-2 Myr and their masses range from 0.5 to 3.5$M_{\odot}$. 
Recently, Ivers et al. (2014) examined mid and far infrared images of NGC 281 at 24, 70, 100, 160 $\micron$ wavelength taken from the Herschel and Spitzer Space Observatories to understand the mechanisms of star formation. They found two different populations of protostars in the region of NGC 281. In the WEST of NGC 281, protostars are found to be within a large molecular cloud obscured by dust while other population of protostars is associated with filamentary pillars and triggered star formation in the East of NGC 281.

We present time series data of stars in the region of the cluster NGC 281. 
The observations of the field containing NGC 281 were initiated in October 2010 and continued till November 2017.
Section 2 describes observations, reduction procedure, comparison of different photometries, variable identification and period determination. Section 3 deals with membership of stars whereas the nature of variable stars identified in the present work has been discussed in section 4. Finally in section 5, we summarize  our results.

\begin{table}
\tiny
%\centering
\caption{Log of the observations. N and Exp. represent number of frames obtained and exposure time, respectively. \label{tab:obsLog}}
\begin{tabular}{lllll}
\hline
S. No.&Date of        &Object&{\it V}                &{\it I}                               \\
         &observations&         &(N$\times$Exp.)&(N$\times$Exp.) \\
\hline
1 & 28 Oct 2010   & NGC 281&4$\times$50s &4$\times$30s \\
2 & 30 Oct 2010  & NGC 281&4$\times$10s, 60s, 200s, 300s&10s,40s, 150s\\
2 & 30 Oct 2010  & SA 98 &2$\times$50s, 15s& 2$\times$30s, 10s, 50s  \\
3 & 28 Nov 2011  & NGC 281&3$\times$50s &-\\
4 & 01 Dec 2011  & NGC 281&80$\times$50s &-\\
5 & 02 Dec 2011  & NGC 281&84$\times$50s &- \\
6 & 06 Dec 2012  & NGC 281&80$\times$50s &- \\
7 & 08 Dec 2012  & NGC 281&51$\times$50s &-\\
8 & 24 Oct 2014   & NGC 281&164$\times$40s &3$\times$50s \\
9 & 25 Oct 2014   & NGC 281&147$\times$40s &-\\
10 & 24 Nov 2014  & NGC 281&112$\times$50s &- \\
11 & 23 Dec 2014  & NGC 281&105$\times$50s &- \\
12 & 27 Oct 2017  & NGC 281&70$\times$40s&-\\
13 & 10 Nov 2017 & NGC 281&139$\times$40s&-\\
\hline
\end{tabular}
\end{table}

\begin{figure}
\includegraphics[width=8cm]{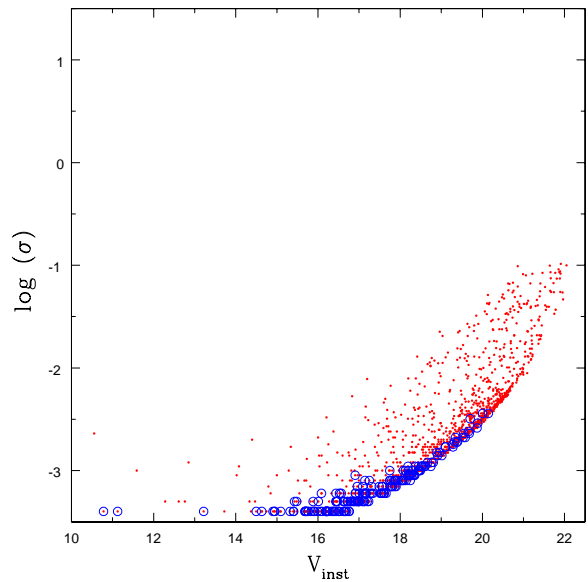}
\caption{Photometric errors as a function of magnitude.
Open circles represent
the variables stars identified in the present work.}
\end{figure}

\begin{figure}
\includegraphics[width=8cm]{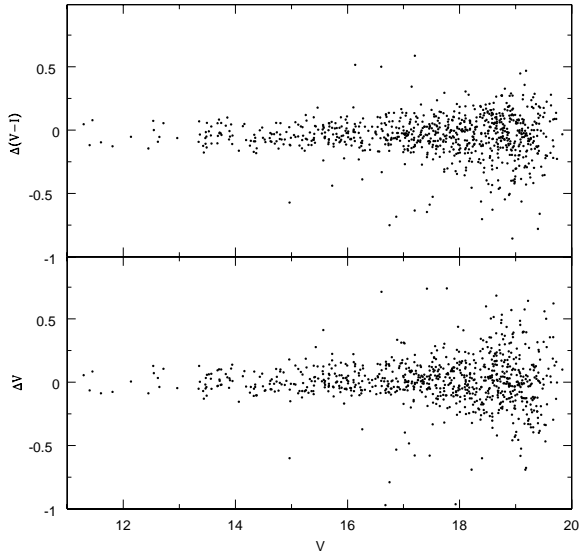}
\caption{Comparison of the present and previous (Sharma et al. 2012). The $\Delta$ represents the difference (present-previous photometry) between the two photometries.}
\end{figure}

\begin{figure}
\includegraphics[width=8cm]{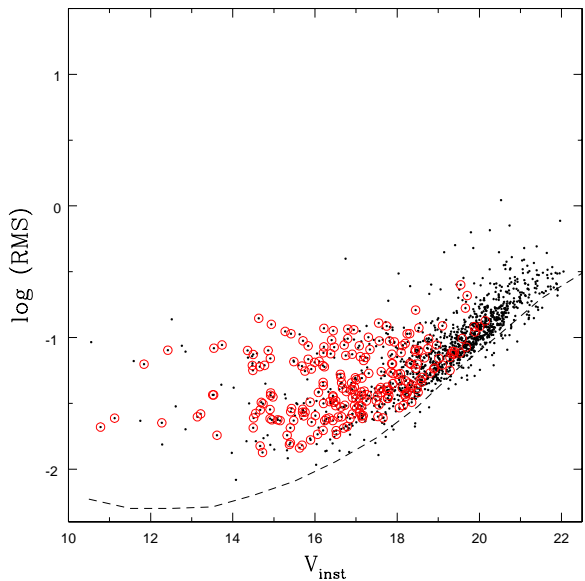}
\caption{Magnitude as a function of standard deviation of each star detected in the present photometry in $V$ band. Open circles represent variable stars identified in the work. The dashed line shows median magnitude scatter expected from the individual standard errors in different magnitude bins. 
}
\end{figure}

\section{Observations and Data Reduction}
Time series $V$ and $I$ band observations of NGC 281 were obtained using a 2048$\times$2048 pixel CCD camera which is
 mounted at the f/13 Cassegrain focus of the 1.04-m ARIES Telescope, Nainital, India. The CCD with a plate scale of 0.37 arcsec/pixel covers a field of  about 13 $\times$ 13 arcmin$^2$ on the sky. The observations were carried out in a binning mode of 2$\times$2 pixels in order to improve the signal-to-noise ratio. During the observations the seeing was around 2 arcsec. 
To standardize the observations, the SA 98 field of Landolt (1992) was also observed. A number of bias and twilight flat frames were also taken during the observations. 

The $V$ band optical observations of NGC 281 were also performed on 27 October 2017 and 10 November 2017 from 1.30-m ARIES Devasthal optical telescope, Nainital, India.
This telescope is equipped with 2048$\times$2048 pixel$^2$ CCD camera with a plate scale 0.54 arcsec per pixel, which gives field of view about 18 arcmin$\times$18 arcmin.

A total of 1046 and 10 observations of the cluster region were secured in $V$ and $I$ bands on 13 and 3 nights, respectively.
The log of observations is given in Table 1.
Fig. 1 shows the observed image in $V$ band from 1.30-m ARIES Devasthal optical telescope.   
The observed images were preprocessed using tasks zerocombine, flatcombine and CCDPROC available in the IRAF\footnote{IRAF is distributed by the National Optical Astronomy Observatory, which is operated by the Association of Universities for Research in Astronomy (AURA) under cooperative agreement with the National Science Foundation} software package. The DAOPHOT package (Stetson 1987) was used to estimate the instrumental magnitudes. To construct the point spread function (PSF) of images isolated stars across the observed field  were selected. 
The PSF photometry of all the sources was obtained using the ALLSTAR task.
DAOMATCH (Stetson 1992) routine of DAOPHOT was used to match up observations taken on various nights with different telescopes. It finds the translation, rotation and scaling solutions between different photometry files, 
whereas DAOMASTER (Stetson 1992) takes .mch file given by DAOMATCH to match the point sources.
DAOMASTER corrects all magnitudes for each star to the magnitude scale of the reference frame.
The instrumental magnitudes were translated to the standard magnitudes using the observations of Landolt SA98 standard stars (Landolt 1992). We derived the transformation
equations which includes the zero points and colour coefficients
as

%**************************************************************
\begin{eqnarray}
v = V + a_{1} - b_{1}\times (V-I) + 0.25\times X   \nonumber\\
i = I+a_{2}  - b_{2}\times (V-I) + 0.10\times X   \nonumber
\end{eqnarray}

where $v$ and $i$ are the  instrumental magnitudes and $X$ is the airmass. The values of $a_{1}$, $a_{2}$, $b_{1}$, and $b_{2}$ are found to be $5.4148645\pm 0.0037297$, 5.6550550$\pm 0.0021983$, $0.0379665\pm 0.0031046$, and $0.0433376\pm 0.0017998$ respectively.

The standard photometric error of the mean magnitude for each star, based on photometric error of individual frame, has been taken 
from the $.mag$ file generated by the DAOMASTER.
The estimated photometric error of all the stars as a function mean instrumental magnitude is shown in Fig. 2.

\begin{figure*}
\hbox{
\includegraphics[width=9.5cm, height=10cm]{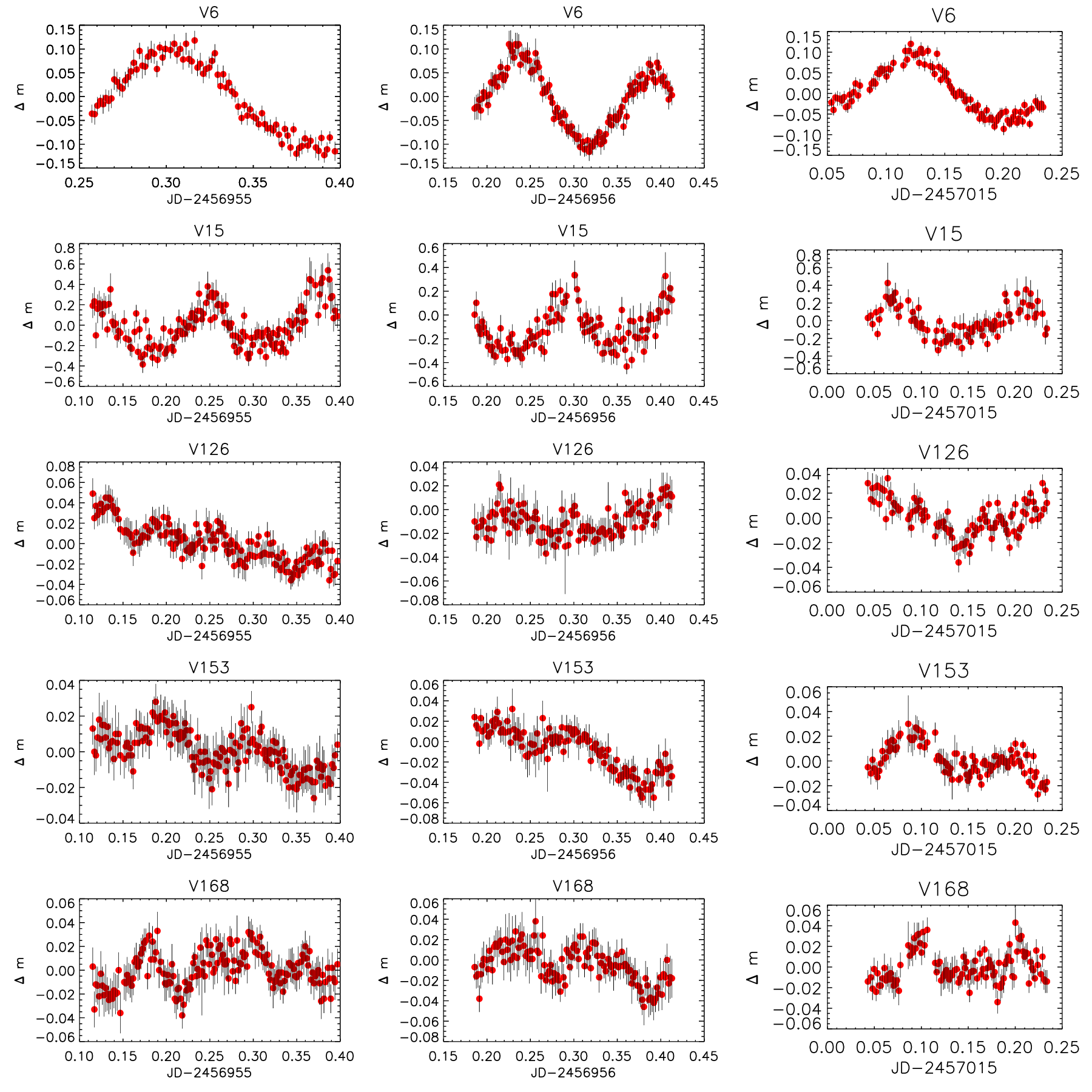}
\includegraphics[width=9.5cm, height=10cm]{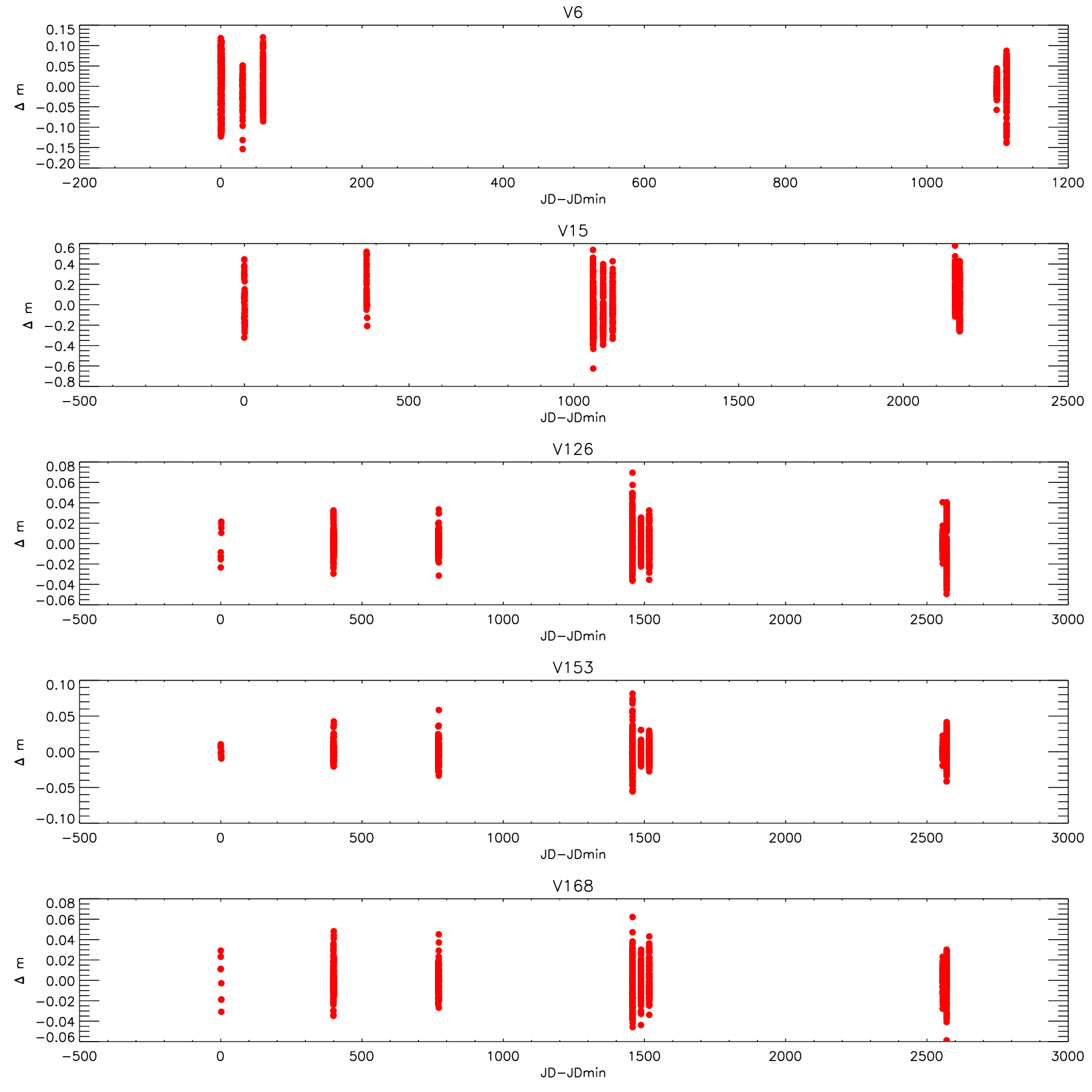}
}
\caption{The $V$ band sample light curves of a few variables identified in the
present work where $\Delta$m represents the differential magnitude. The individual night observations of stars are shown in left panel 
while right panel represents all the observations of stars. The complete Figure 5 is available in electronic form.}

\end{figure*}
%*************************************************************************************************************************************************

\subsection{Comparison with Previous Photometry}
The present CCD photometric data have been compared with the CCD observations given by Sharma et al. (2012).
We have found 867 common stars between these two photometries.
The difference $\Delta$ (present data -literature data) as a function of $V$ magnitude is shown in Fig. 3, which
indicates that the present $V$ magnitudes are in agreement with those given by Sharma et al. (2012), whereas a systematic difference of 0.05 mag in $(V-I)$ colour is present between the two photometries. 

\begin{figure*}
\hbox{
\includegraphics[width=9cm, height=9cm]{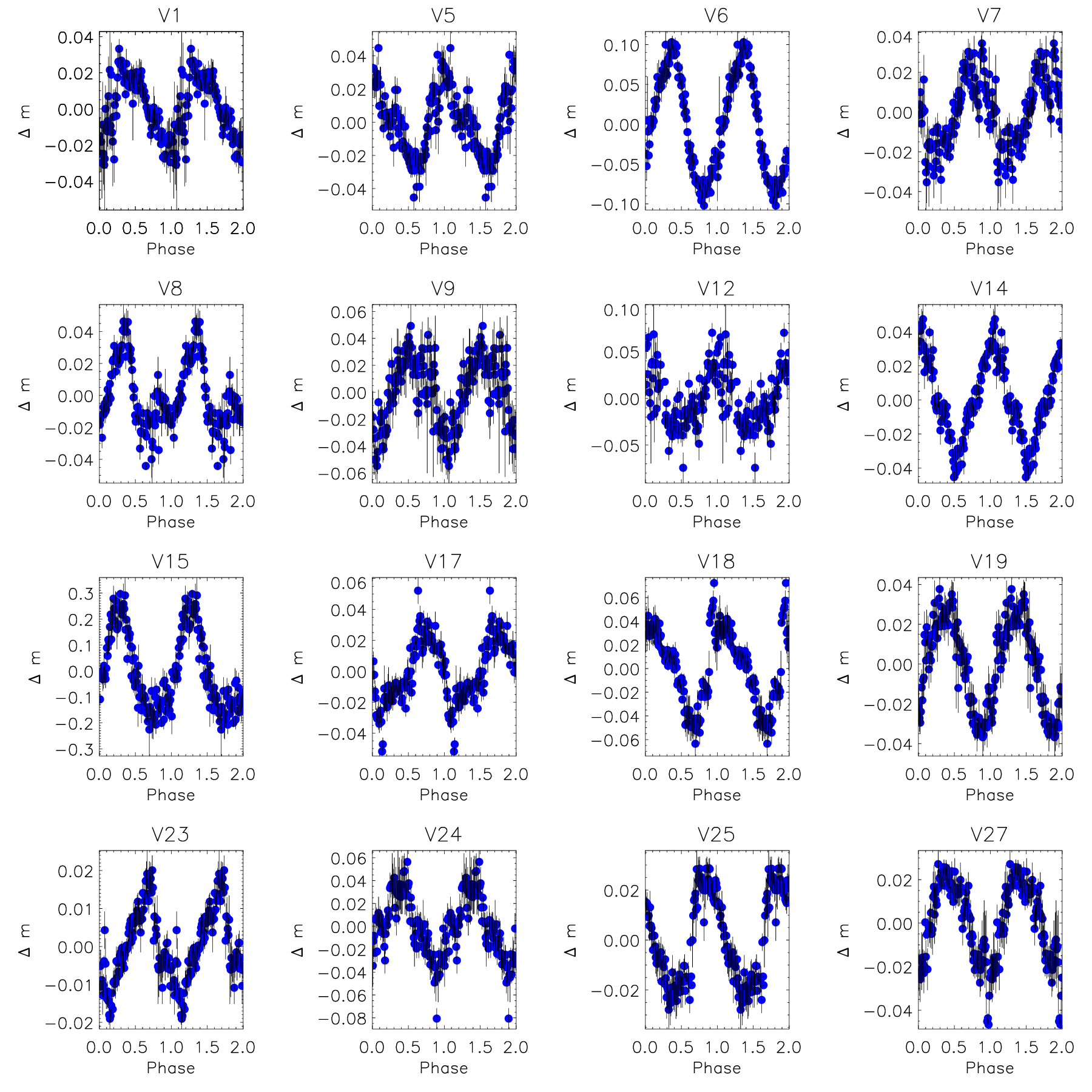}
\includegraphics[width=9cm, height=9cm]{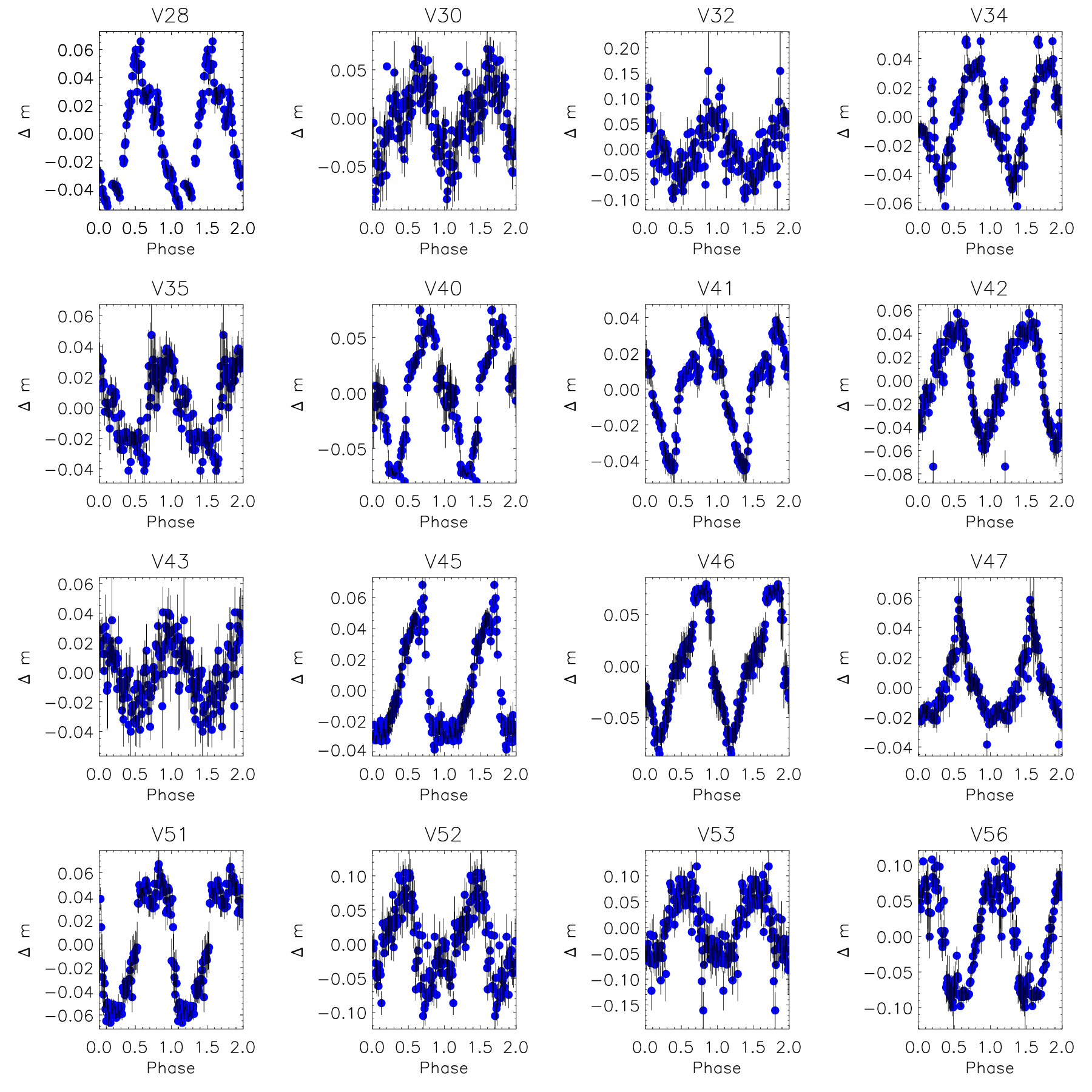}
}
\hbox{
\includegraphics[width=9cm, height=9cm]{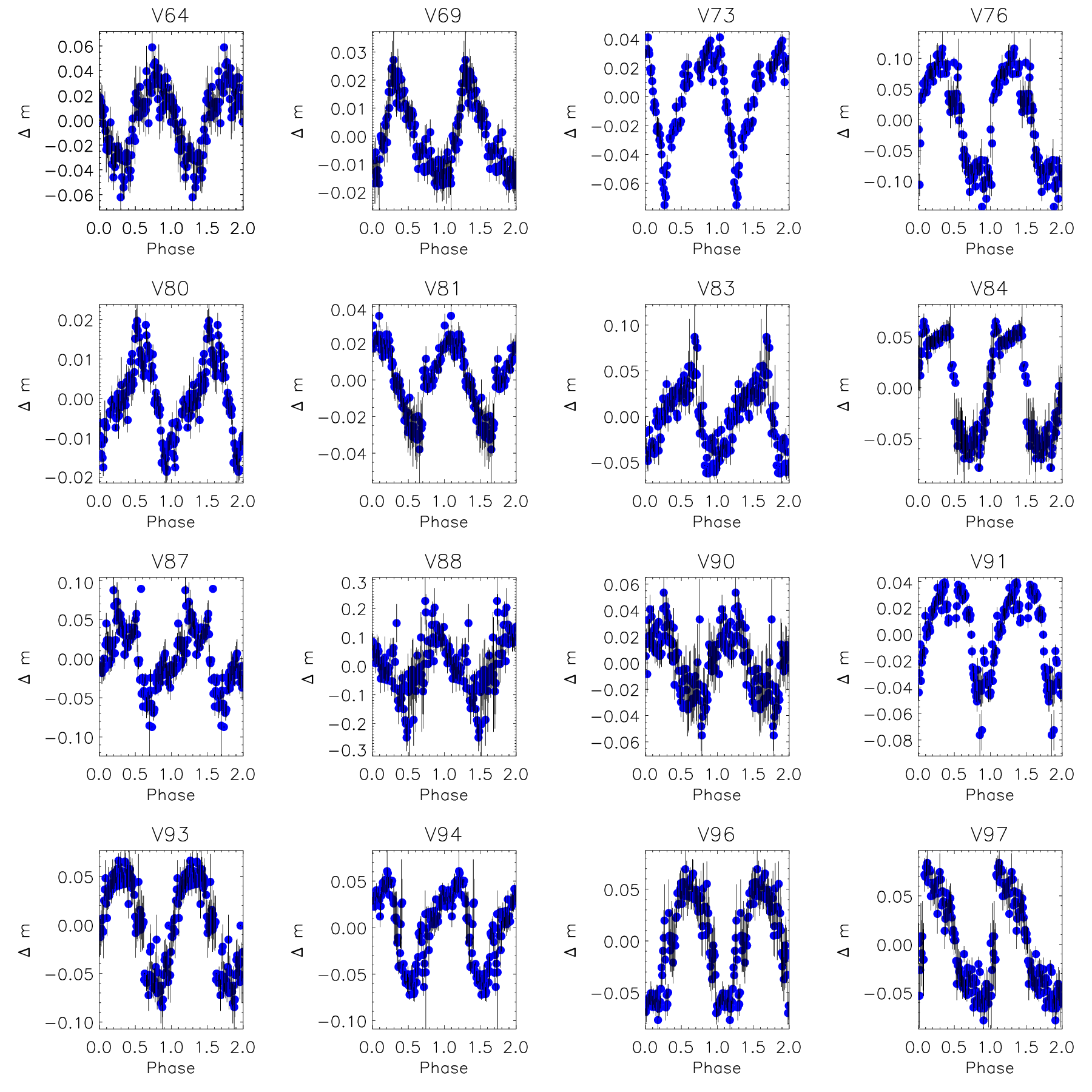}
\includegraphics[width=9cm, height=9cm]{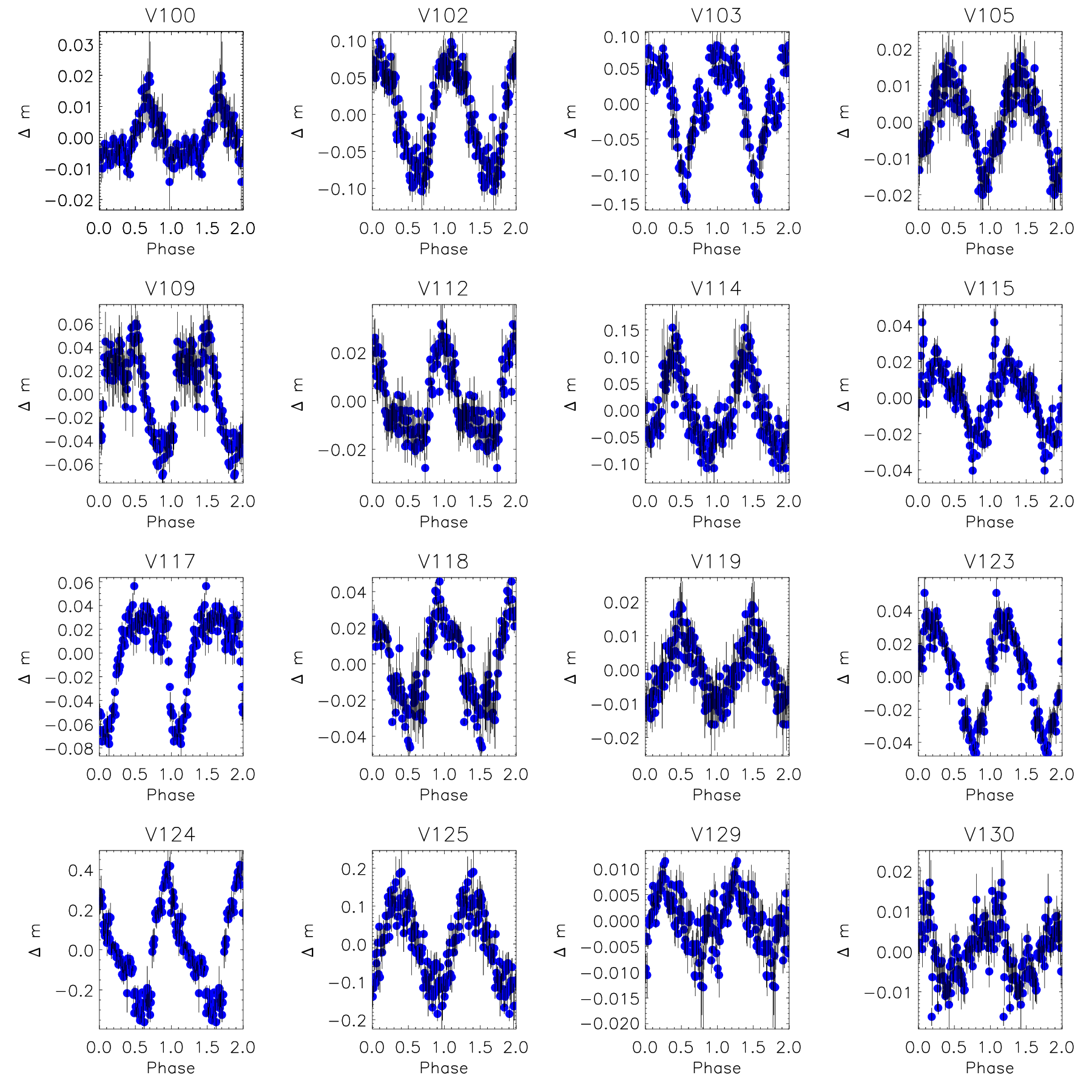}
}
\caption{The $V$ band phased light curves of periodic variable stars. }
\end{figure*}

\setcounter{figure}{5}
\begin{figure*}
\hbox{
\includegraphics[width=9cm, height=9cm]{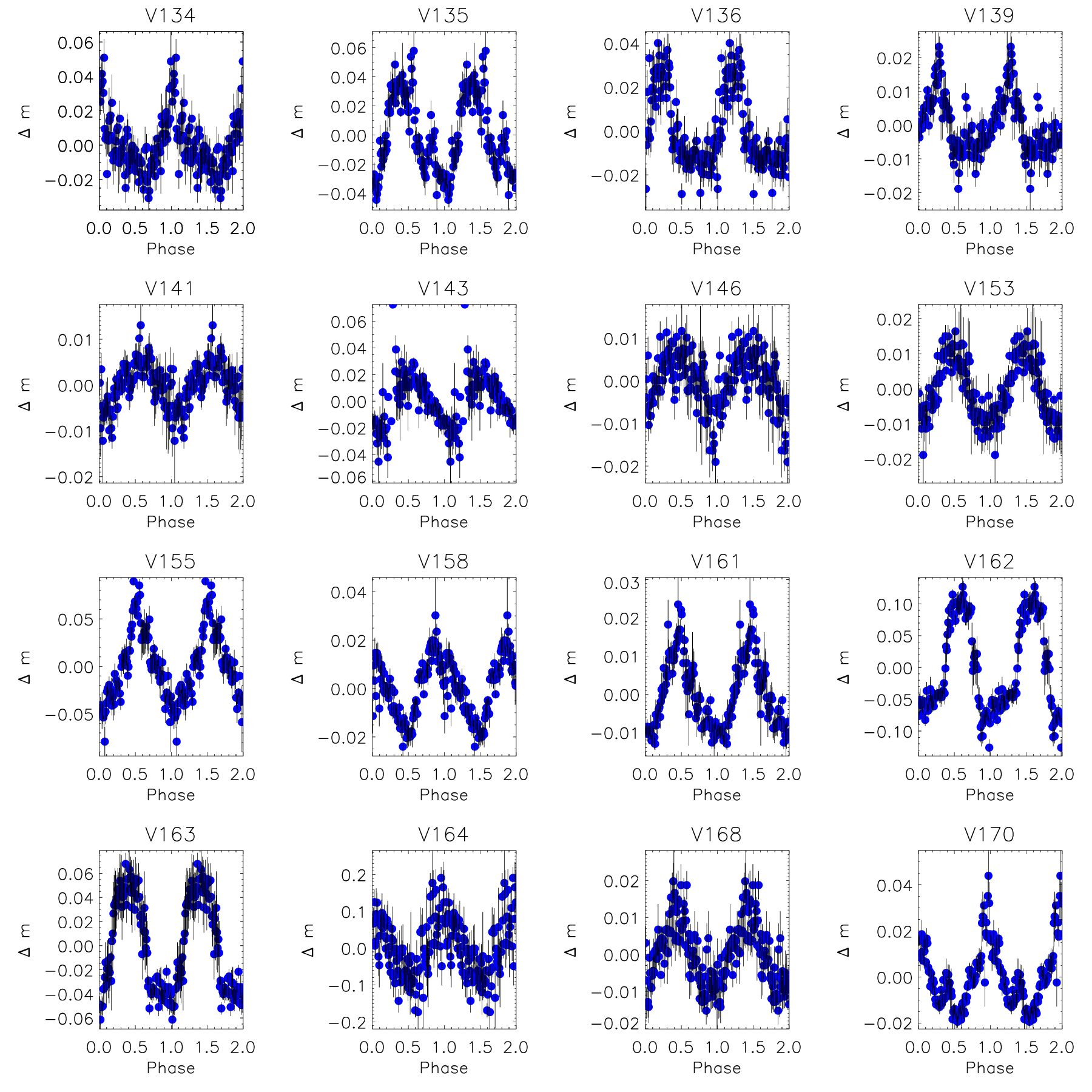}
\includegraphics[width=9cm, height=9cm]{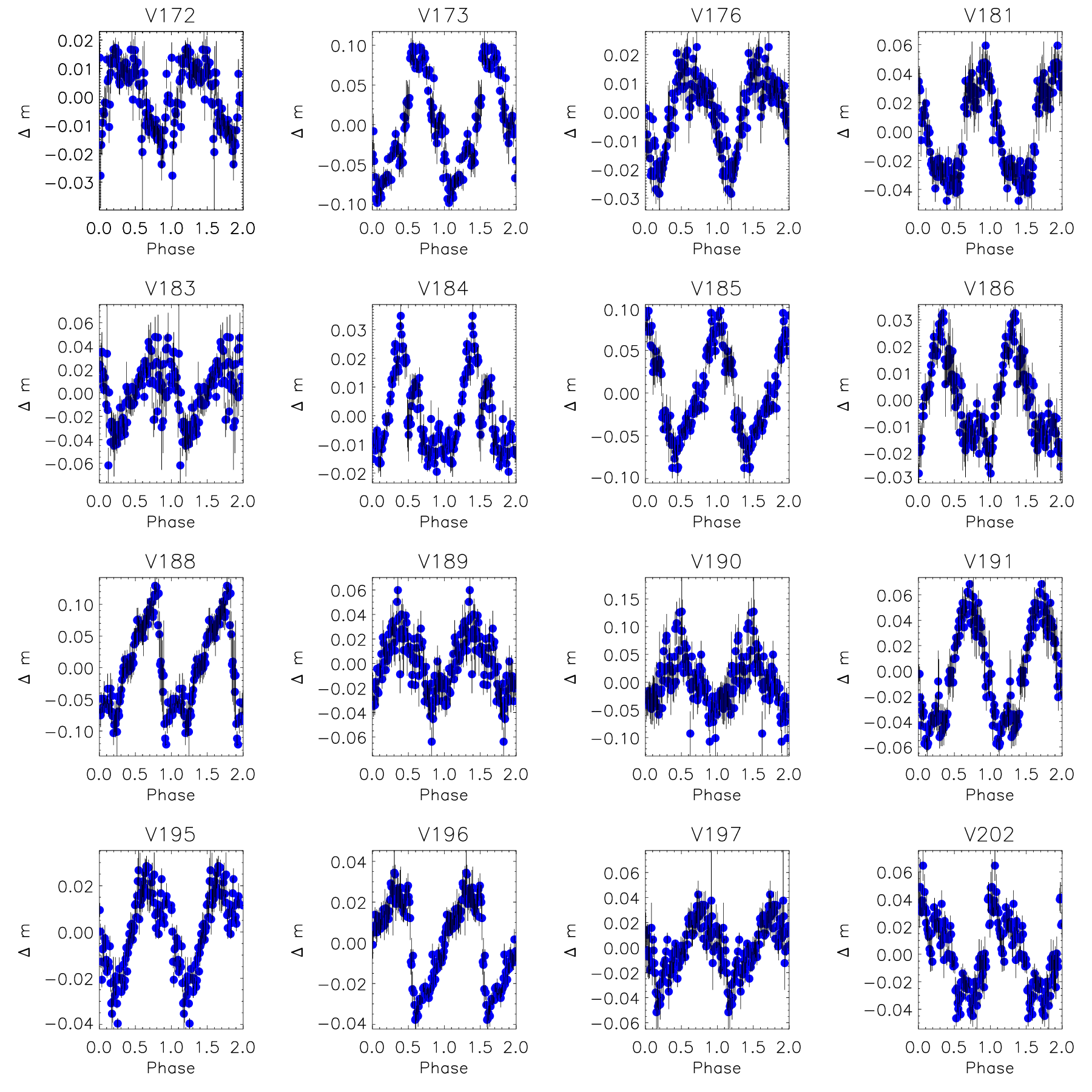}
}
\hbox{
\includegraphics[width=9cm, height=9cm]{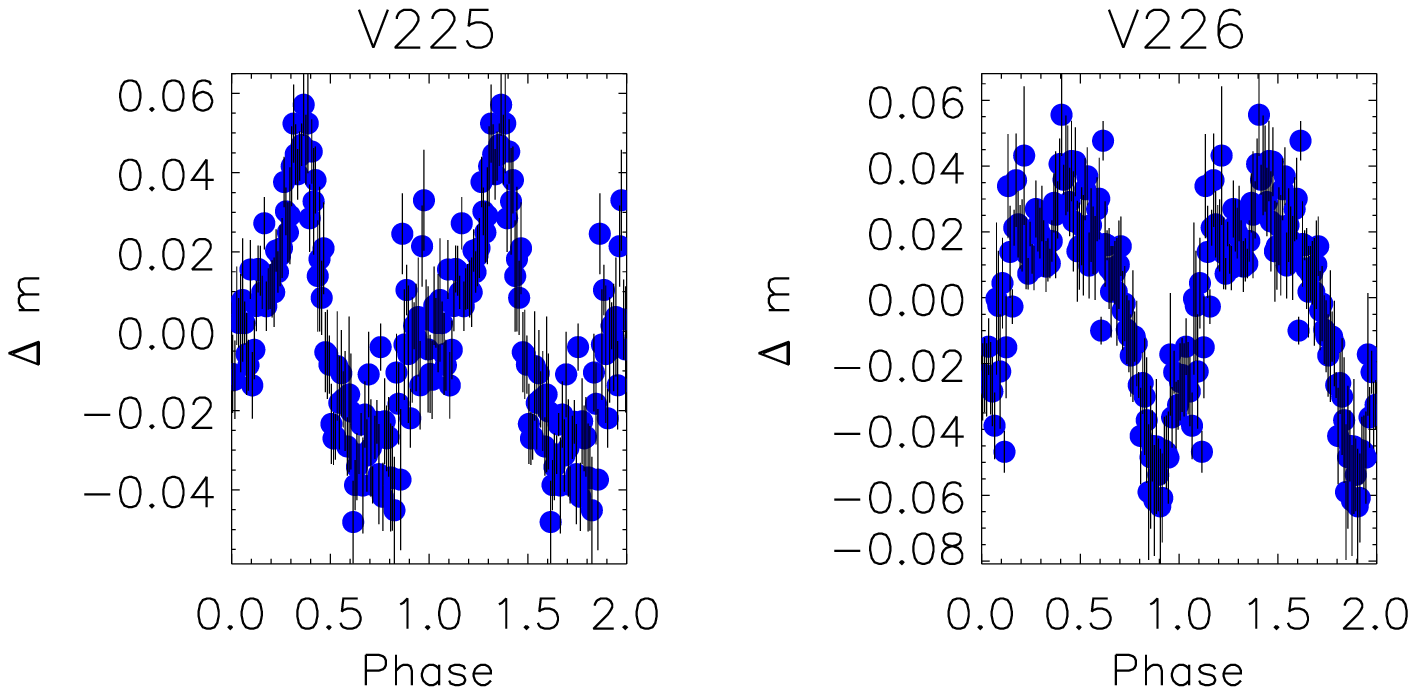}
}
\caption{Continued. }
\end{figure*}

\setcounter{figure}{6}
\begin{figure*}
\hbox{
\includegraphics[width=9cm, height=9cm]{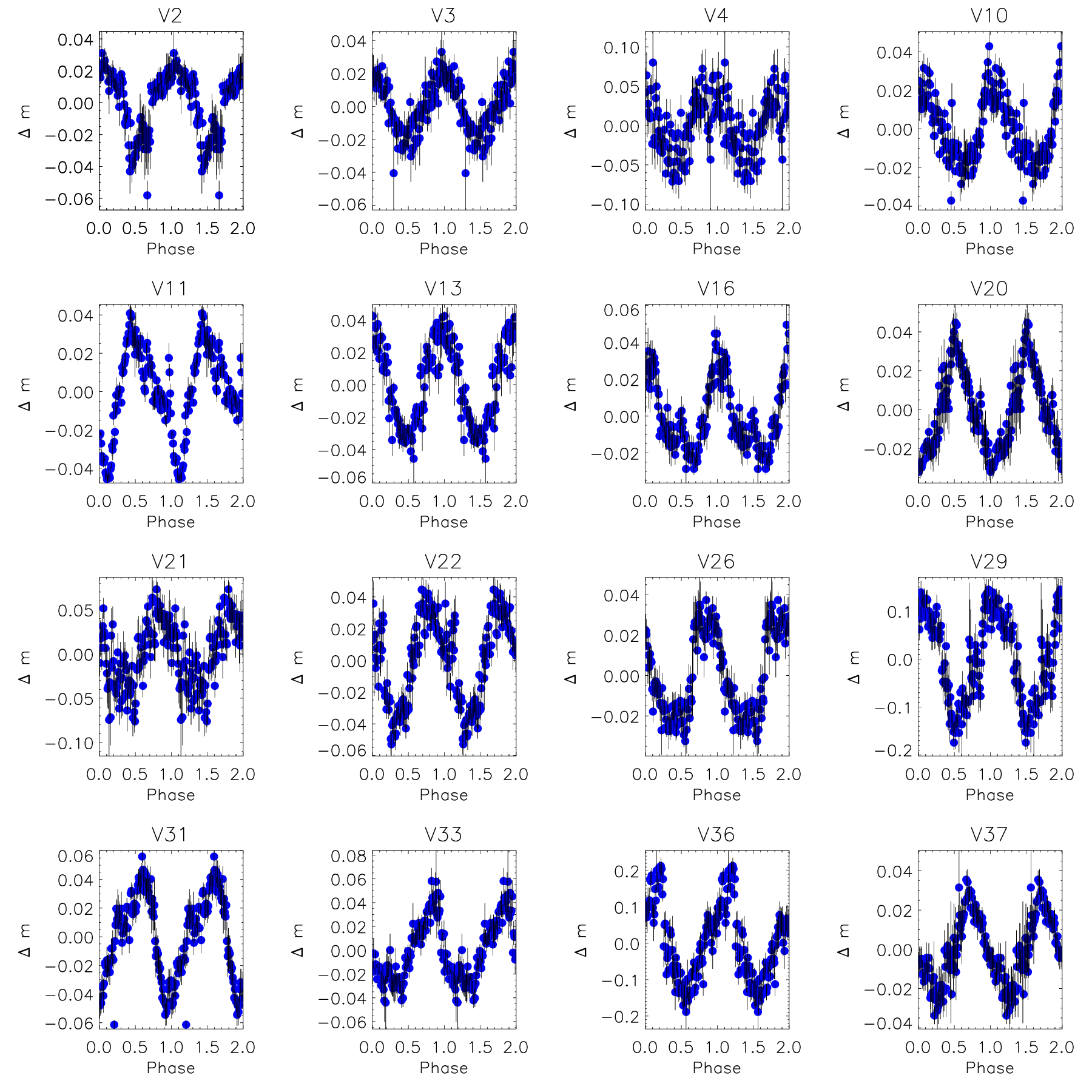}
\includegraphics[width=9cm, height=9cm]{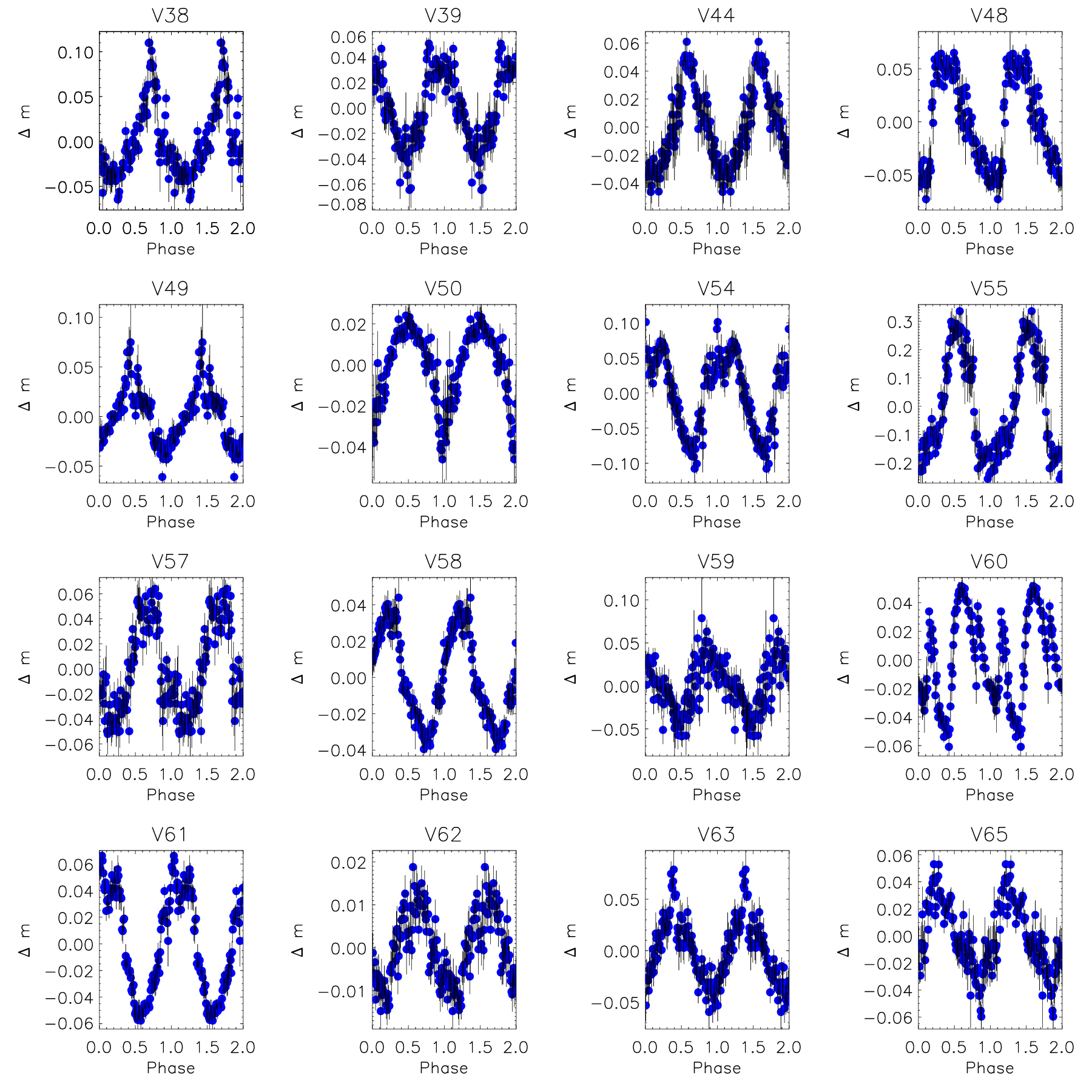}
}
\hbox{
\includegraphics[width=9cm, height=9cm]{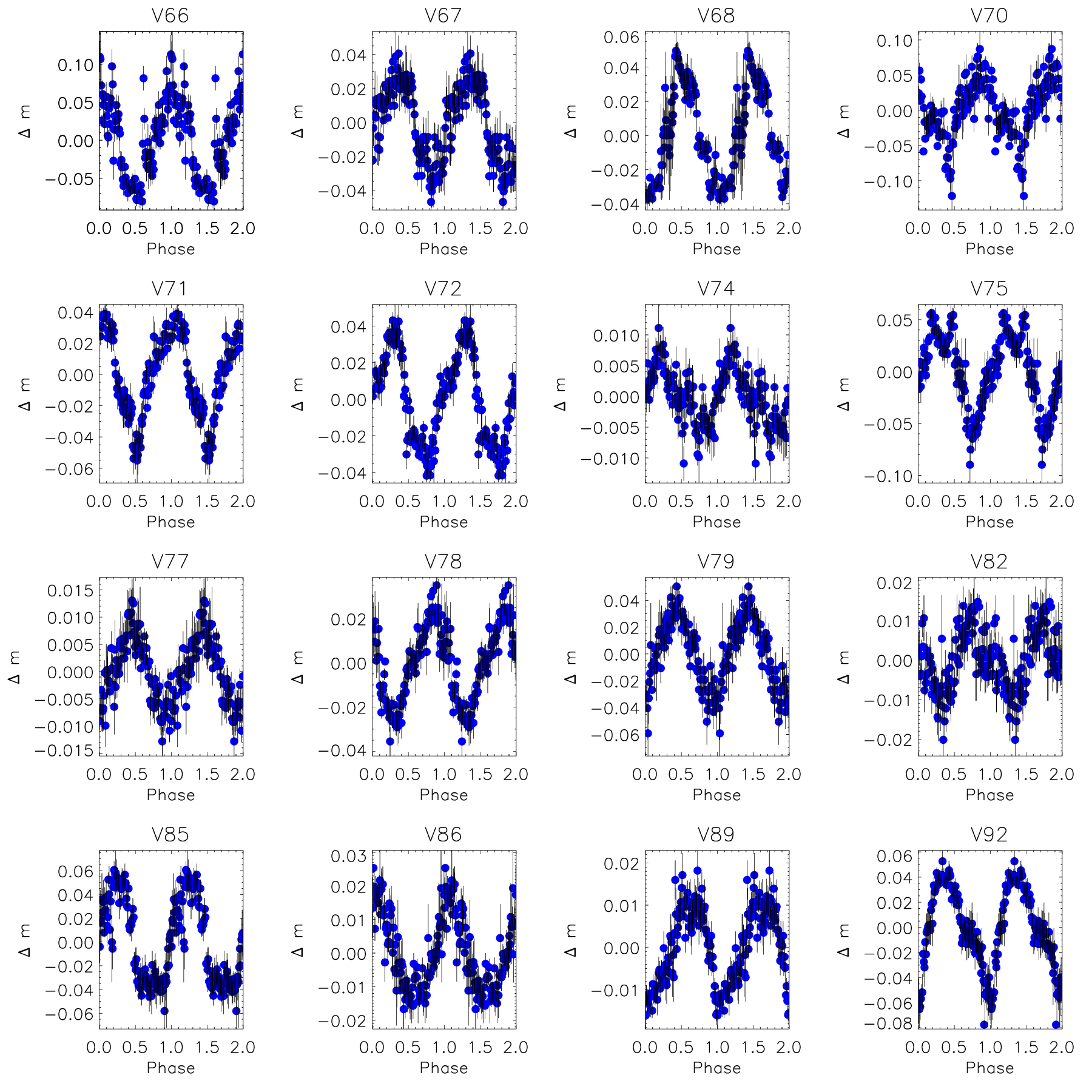}
\includegraphics[width=9cm, height=9cm]{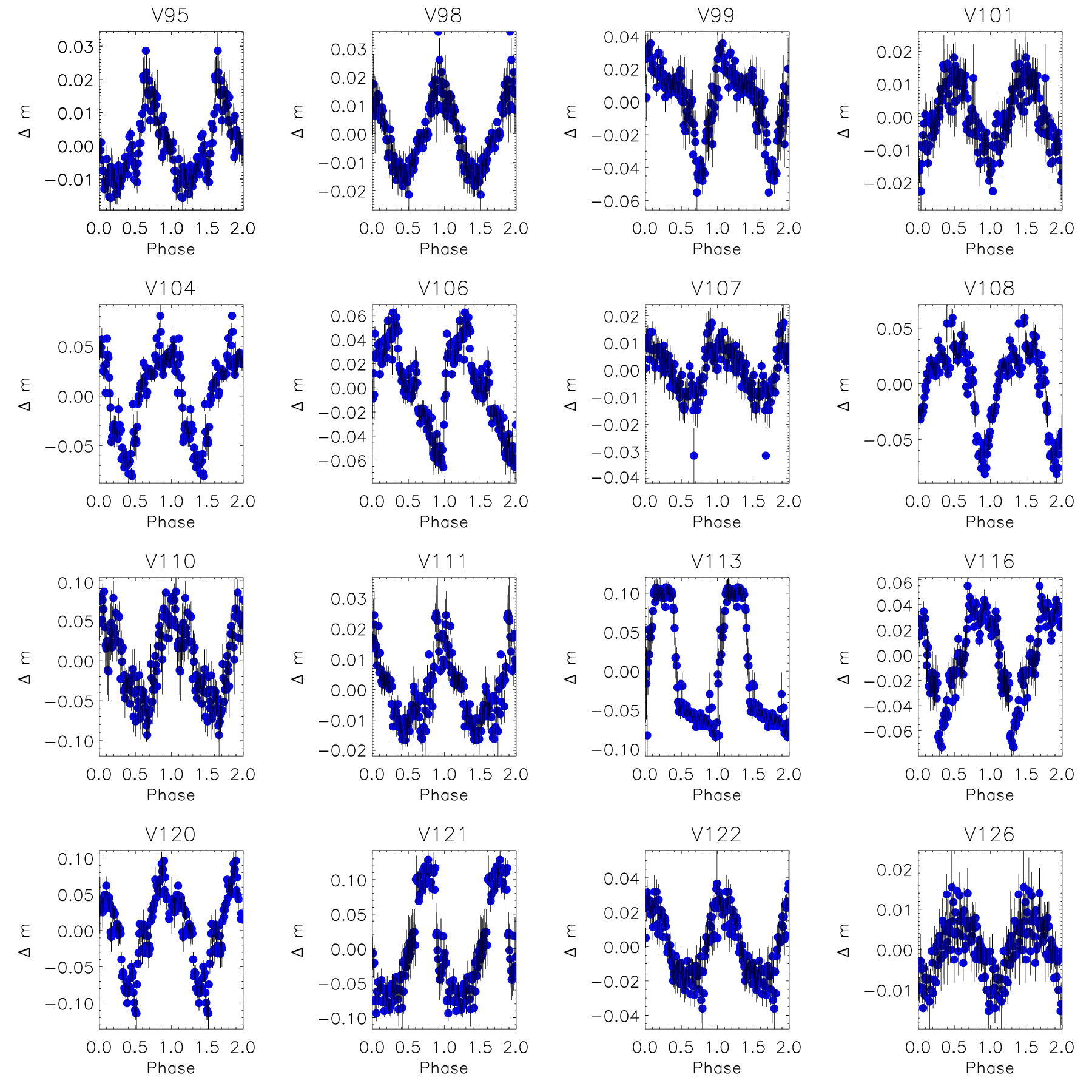}
}
\caption{The $V$ band phased light curves of probable periodic variable stars. }
\end{figure*}

\setcounter{figure}{6}
\begin{figure*}
\hbox{
\includegraphics[width=9cm, height=9cm]{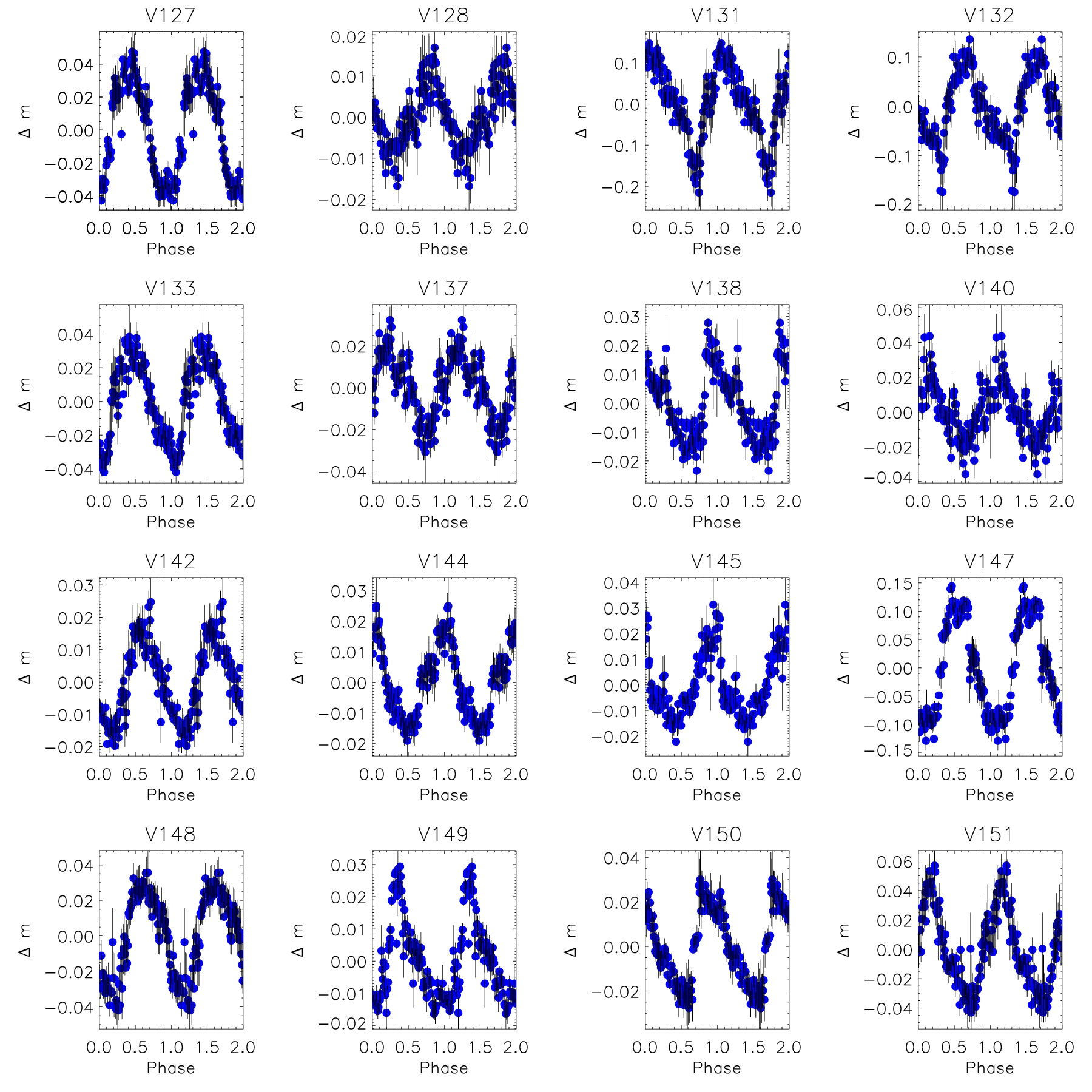}
\includegraphics[width=9cm, height=9cm]{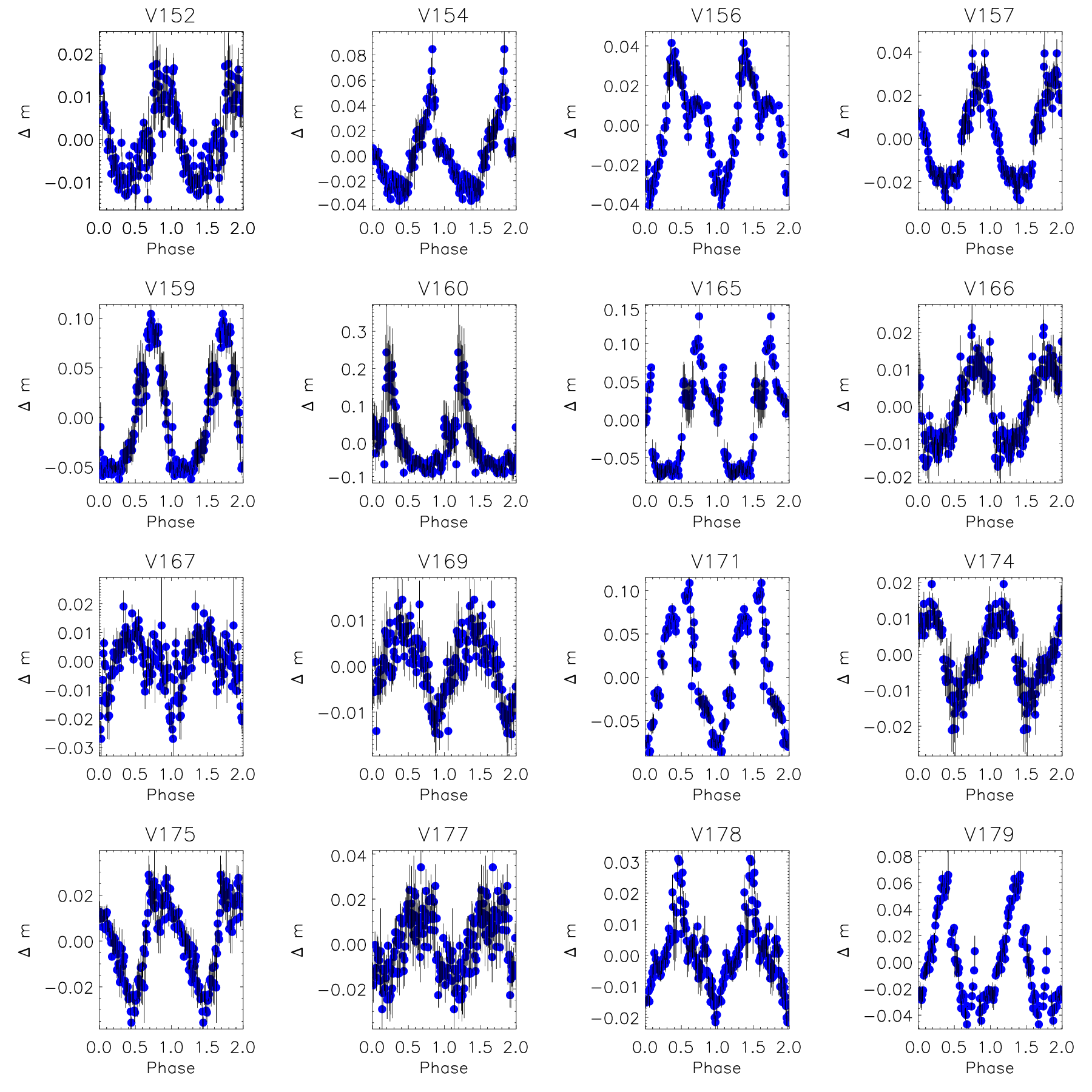}
}
\hbox{
\includegraphics[width=9cm, height=9cm]{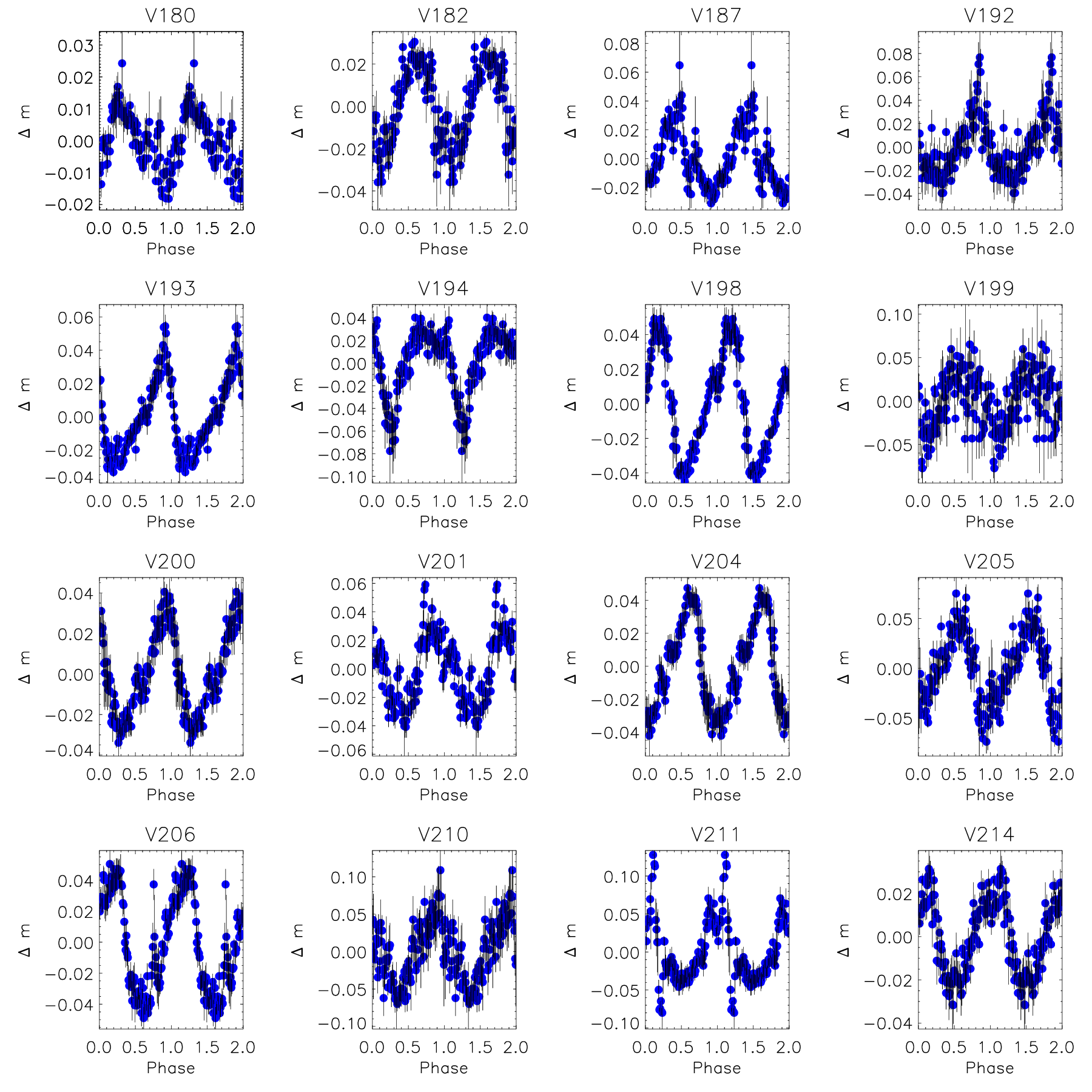}
\includegraphics[width=9cm, height=9cm]{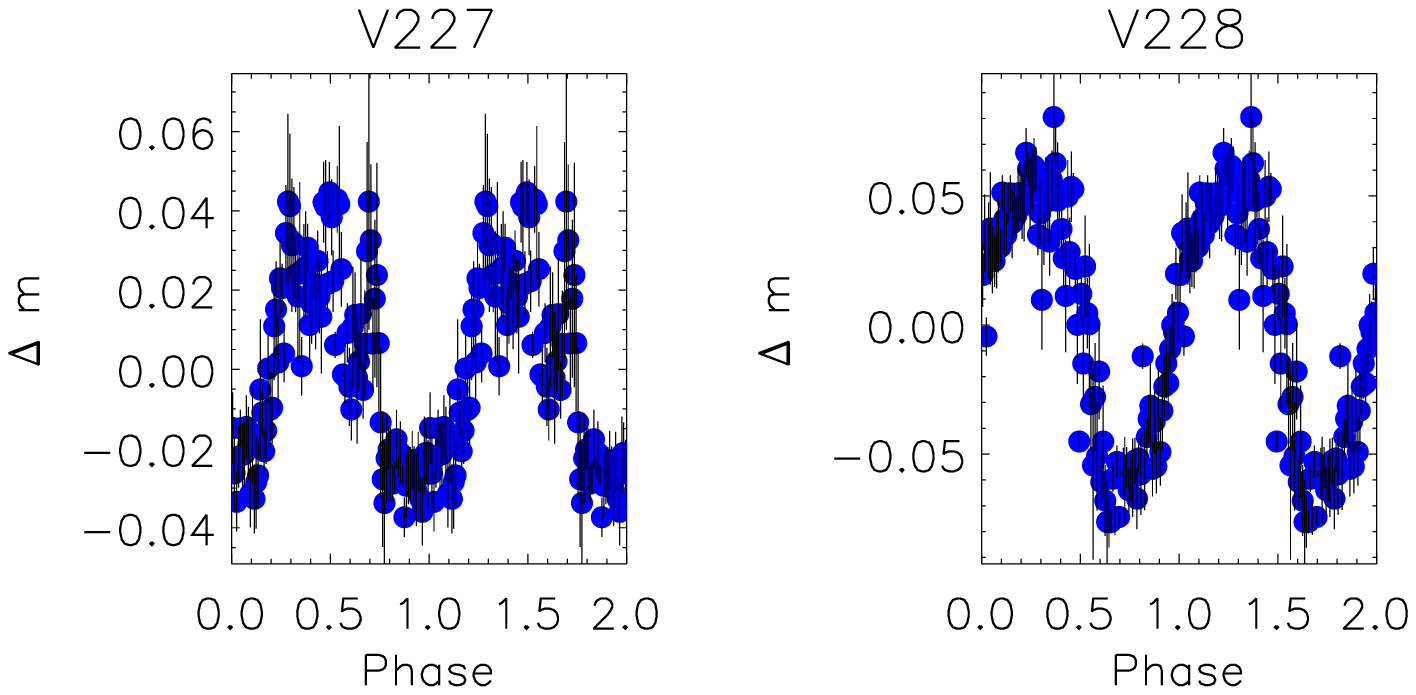}
}
\caption{Continued. }
\end{figure*}

\subsection{Variables identification}
The differential magnitudes ($\Delta$m) in the sense variable minus comparison star were plotted against  Julian date (JD).
To obtain differential magnitude of target star (V), we have selected three non variable stars C1, C2 and C3 as comparison stars in the field of NGC 281. Then,
magnitude difference V-C1, C1-C2, C1-C3 have been determined relative to C1. The comparison stars C2 and C3 have been used to check non variable nature of C1.  
 The probable variables were identified visually by inspecting the light curves. The visual inspection yielded  228 variable candidates. 
In order to verify variability of these stars, $rms$ (root mean square) scatter is computed for each star. 
The observed $rms$ scatter includes both the intrinsic variability and the mean photometric error. 
In Fig. 4 we have plotted $rms$ of each observed star as a function of magnitude, where open circles show the variables identified in the present work.
Fig. 4 indicates that generally identified variables have higher $rms$ value. The lower 
$rms$ stars in general are low amplitude variables.
There are 4 variables with $rms$ $>$ 3 times the mean $rms$ of the magnitude bin. 
Ten and 48 variables have $rms$ $>$ 2 times the mean rms of the magnitude bin and $rms$ $>$ 1 times the mean $rms$ of the magnitude bin, respectively.
The light curves of some of the stars having large $rms$ do not show any
variability.  
These stars are either having some outliers or located on edge of CCD frame.   
Low amplitude variables have been detected visually by inspecting their light curves.
The photometric variability of stars identified as variables has also been checked 
using the $Chi$ square test (Sesar et al. 2007). All 228 variables have probability $\ge$ 90\% except four stars V129, V136, V139, V141 and 
V146 for which probability to be a variable is less than 90\%.
We considered these stars as variable on the basis of visual inspection of their light curves.
The sample light curves of a few variables are shown in Fig 5.
The identification number, coordinates and optical data for these variable stars are given in Table 2. The CCD pixel coordinates of these identified variables were converted to celestial coordinates (RA and DEC) for J2000 with the help of the CCMAP and CCTRAN tasks in IRAF.
The reference astrometric catalog was made from DSS2 R band image (https://skyview.gsfc.nasa.gov/current/cgi/query.pl) which was used in CCMAP to determine the plate solution for the image.
The variable candidates identified in the present work are marked in Fig. 1.

To determine periods of variable stars we applied the Lomb-Scargle (LS) periodogram (Lomb 1976; Scargle 1982). 
This periodogram is known to work well if data are unevenly sampled.
 The false alarm probabilities corresponding to the power of maximum frequency  
from LS for all the identified variables come out to be zero. The peak of the 
periodogram  ranges from 0.076 to 0.92 for variable stars. The range of highest periodogram peak for the present variables is between $\sim$0.021 to $\sim$0.038, $\sim$0.023 to $\sim$0.040 and $\sim$0.026 to $\sim$0.046 at 10\%, 5\% and 1\% false alarm probabilities, respectively.  
The light curves of stars were folded with calculated period. We visually inspected the phased light curves and opted for the period showing the best phased light curve. The most probable periods with amplitude are listed in Table 2.
The phased light curves of periodic and semi periodic/irregular variables candidates are 
shown in Fig. 6 and Fig. 7, respectively where averaged differential magnitude in 0.02 phase bin along with $\sigma$ error bar has been plotted. A few of 
the individual data points which show relatively large photometric errors have not been considered in folding the light curves.
During calculation of period for short period variable stars, if there was a vertical shift between the observations for different nights, 
it was calculated and applied to bring different night observations at same level.
In the present study several stars seem to have many        
periods. These could be semi periodic/irregular variables (light curves shown in Fig. 7).
The
light curves of these stars were folded with their best periods as folded data with the best period gives smooth phased light curve of the star. 
These irregular/semiperiodic variables could be binary systems with pulsating components with multiple periods or
other complex systems. 
From now, we have considered them as probable periodic variable stars.
The phased light curves will be further discussed in Section 7.

\section{Cluster membership of the identified variables}
The characterization of the identified variables needs information whether these are the members of the cluster or not. 
The $UBVI$ ($UBV$ data taken Sharma et al. 2012 and present $VI$ data) and $JHK$ photometric observations of the 
cluster NGC 281 (Cutri et al. 2003) have been used to establish the membership of the
identified variables. In addition, the membership of a star is also verified with the help of proper motion data taken from Gaia 
astrometric mission (Gaia Collaboration et al. 2018) 

\subsection{U-B/B-V Two colour diagram (TCD)} 
The $U-B/B-V$ TCD is a useful tool to identify MS members of the cluster region. The $U-B$ data 
for only 128 identified variables are available in Sharma et al. (2012). Fig. 8 plots $U-B/B-V$ TCD for 128 variables 
where the continuous curves show ZAMS from Girardi et al. (2002), which are shifted along the reddening vector having a slope of $E(U-B)/E(B-V)$ = 0.72. The distribution of MS variables reveals a variable reddening in the cluster region with a minimum reddening $E(B-V)$ = 0.32 mag. The sources lying within the MS band having $E(B-V)$ = 0.32 mag to 0.45 mag with O to A spectral type can be considered as possible MS members of the cluster. The 
identification as well as estimation of reddening for YSOs is not possible using the $U-B/B-V$ TCD because the $U$ and $B$ band fluxes may be affected by excess due to accretion. Probable MS cluster members are mentioned in Table 2.

\subsection{J-H/H-K TCD}
Since young stellar objects (YSOs) generally show $H{\alpha}$ emission, NIR excess or X-ray emission, therefore $J-H/H-K$ TCD is one of the very useful tools to identify PMS objects. Fig. 9 displays $J-H/H-K$ TCD for all the identified variables. $JHK$ data have been taken from the 2MASS catalogue (Cutri et al. 2003) which was further converted to CIT system 
using the relations given in the 2MASS website (http://www.astro.caltech.edu/ jmc/2mass/v3/transformations/). In Fig. 9, the sources lying in `F’ 
region could be either field stars or Class III (WTTSs) and Class II (CTTSs) sources with small NIR excesses. The sources lying in the `T' region can be considered mostly as Class II objects/ CTTSs (for detail, Lata et al. 2019).
A comparison of the TCD of the NGC 281 region with the NIR TCD of nearby reference region (figure 6 of Sharma et al. 2012) indicates that the sources lying in the `F’ region above the extension of the intrinsic CTTS locus as well as sources having $(J-H)$ $\gtrsim$ 0.6 mag and lying to the left of the first (left-most) reddening vector could be WTTSs/Class III
sources. Sharma et al. (2012) have also identified YSOs on the basis of $H{\alpha}$ emission, NIR TCD, MIR TCD, and X-ray emission.
A comparison of present data with that by Sharma et al. (2012) yields 11 common PMS stars. 
The MIR observations are very useful, which provide information of YSOs remained embedded in the molecular clouds. 
Several studies (Getman et al. 2012; Jose et al. 2013) classify YSOs with help of MIR TCDs. In the IRAC colour plane, YSOs occupy distinct regions according to their nature. 
The figure 7 of Sharma et al. (2012) presents a [5.8]-[8.0] versus [3.6]-[4.5] TCD for the observed sources. This figure shows that class II, III and 0/I
sources have different locations in these TCDs. 
The common YSOs are V76, V102, V103, V114, V133, V152, V156, V164, V171, V188 and V209.  In $J-H/H-K$ TCD star V152 lies below the 
intrinsic locus of T Tauri stars and right of middle reddening vector. 
Out of these 11 stars two sources V133 and V156 are found to be lying in the location of Herbig Ae/Be stars in $J-H/H-K$ TCD.  
Details of the probable PMS cluster members are mentioned in Table 2.

\subsection{V/V-I CMD}
The $V/V-I$ CMD for 225 identified variables is displayed in Fig. 10. The $V-I$ colours for three variables were not available. The zero-age main-sequence (ZAMS) 
by Girardi et al. (2002) as well as PMS isochrones for various ages and evolutionary tracks for various masses by Siess et al. (2000) have also been plotted. 
Assuming the reddening $E(V-I)$ towards the cluster as 0.40 mag, the comparison of observations with the ZAMS by Girardi et al. (2002) yields 
a distance modulus of 13.25 mag, which corresponds a distance of 2.82 kpc, which is in good agreement with the value (2.81 kpc) used by Sharma et al. (2008). 
The probable MS, PMS variables identified on the basis of CMD are shown with triangle and filled square symbols in Fig. 10, respectively. The variables having membership 
probabilities $\ge$ 50\% (see next section) are encircled. 
Based on the above mentioned TCDs and CMD, we have established membership of 81 stars, of which 30 and 51 
stars could be probable MS and PMS variables, respectively. Remaining 147 variables may belong to the field region. The classification of variables is given in the last column of Table 2.
\begin{figure}
\includegraphics[width=9cm]{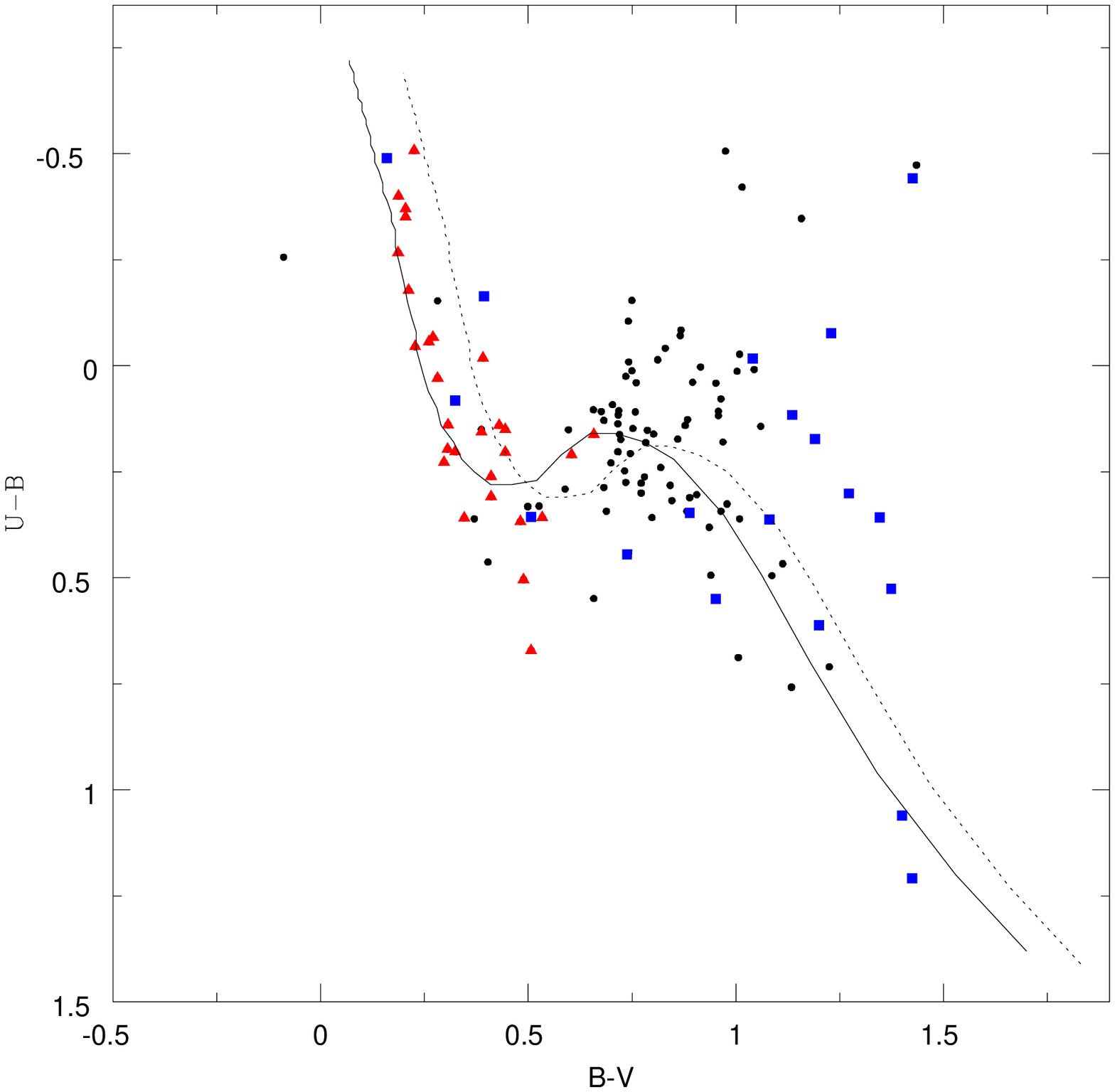}
\caption{$(U-B)/(B-V)$ TCD for variable stars identified in the present study. All the $UBV$ data are taken from
Sharma et al. (2012). The continuous and dotted line represent the ZAMS (Girardi et al. 2002) which are shifted along the reddening vector for reddening $E(B-V)= 0.32$ mag and 0.45 mag.
 Triangles are identified as MS variables and filled squares show PMS variables.
}
\end{figure}

\begin{figure}
\includegraphics[width=9cm]{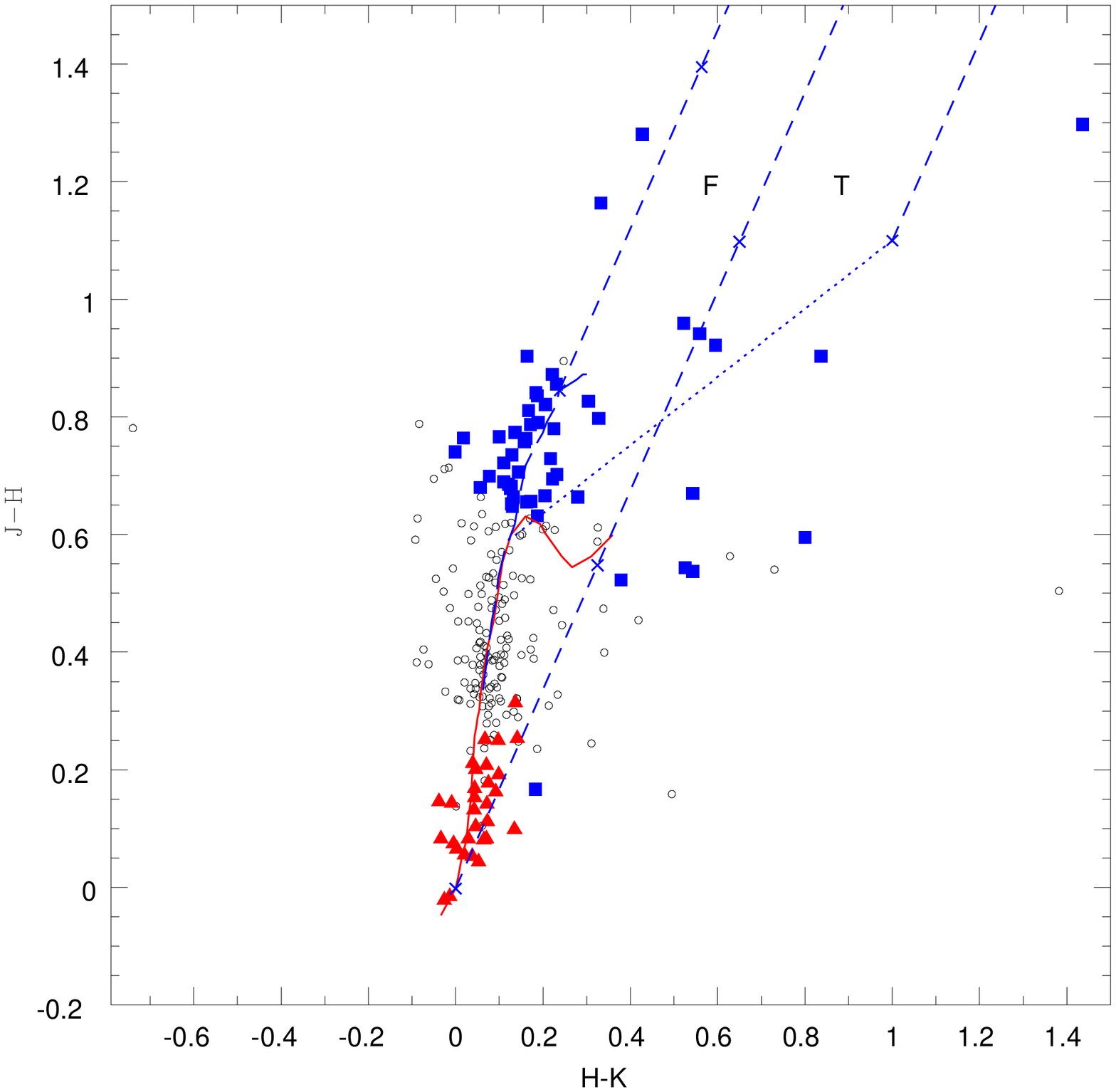}
\caption{$(J-H)/(H-K)$ TCD for variable stars detected in the field of
 NGC 281. The $JHK$ data have been taken from the 2MASS catalogue (Cutri et al. 2003). The continuous and long dashed line show sequences for dwarfs and giants (Bessell \& Brett 1988), respectively. The TTSs locus (Meyer et al. 1997) is shown by dotted line.
The small dashed lines are reddening vectors (Cohen et al. 1981) and
an increment of visual extinction of $A_{V}$ = 5 mag is denoted by crosses on the reddening vectors.
Filled squares and triangles represent PMS and MS whereas open circles may be either MS members of the cluster or field stars.
The `F' and `T' regions are explained in Section 3.2.
}
\end{figure}

\begin{figure}
\includegraphics[width=9cm]{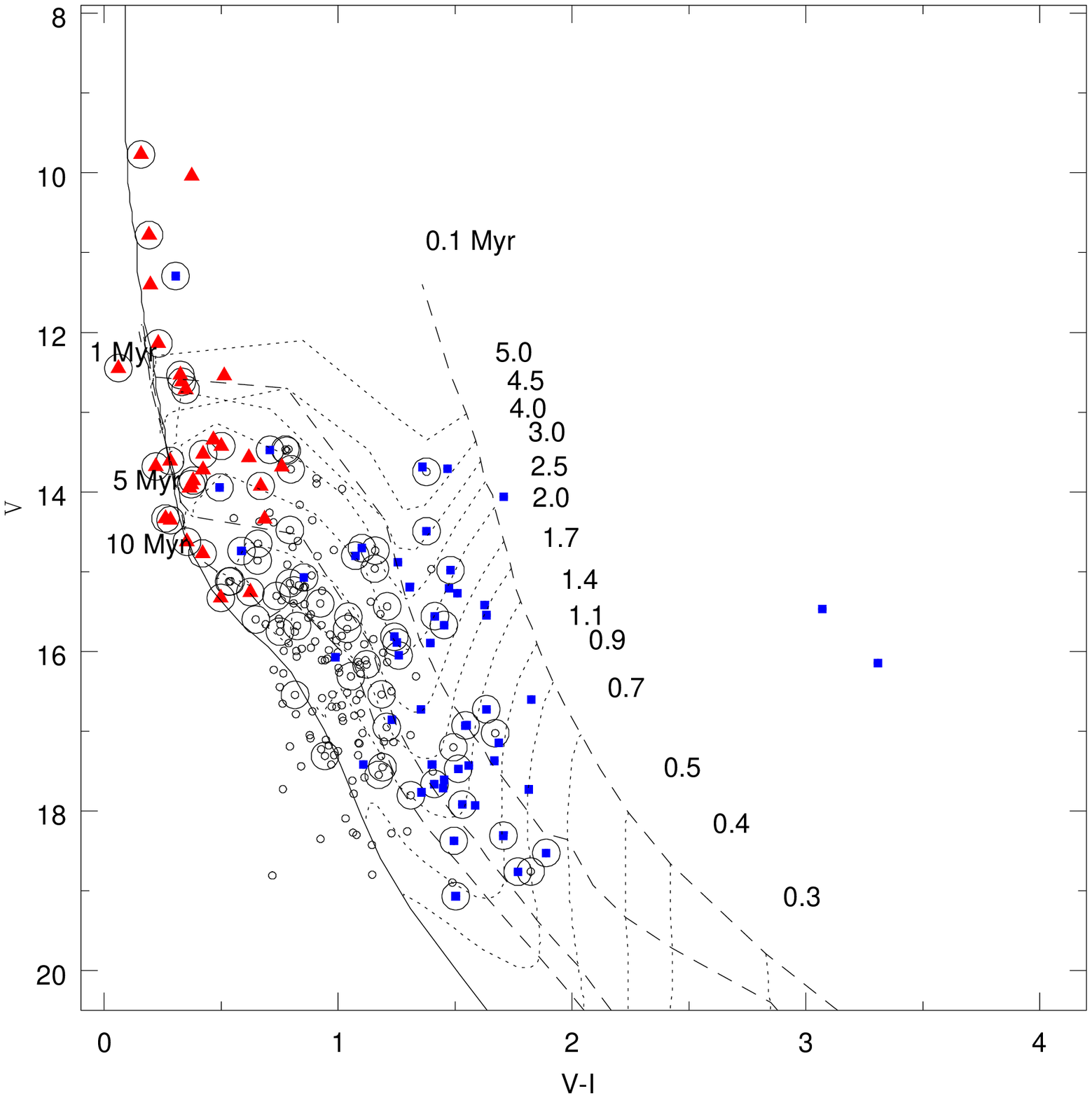}
\caption{$V/(V-I)$ CMD for variable stars in the region of the cluster NGC 281.
The filled squares represent probable PMS variable stars, whereas triangles show MS stars. The variable stars belong to the field population is shown by open circles.
The encircled points here refer to those stars which are having membership probability $\ge$50\%.
The continuous curve shows ZAMS by Girardi et al. (2002) while dashed lines
 represent PMS isochrones for
0.1, 1, 5, 10 Myrs (Siess et al. 2000).
The PMS evolutionary tracks for different masses taken from Siess et al. (2000) are shown by dotted curves.
}
\end{figure}

\begin{figure}
\includegraphics[width=9cm]{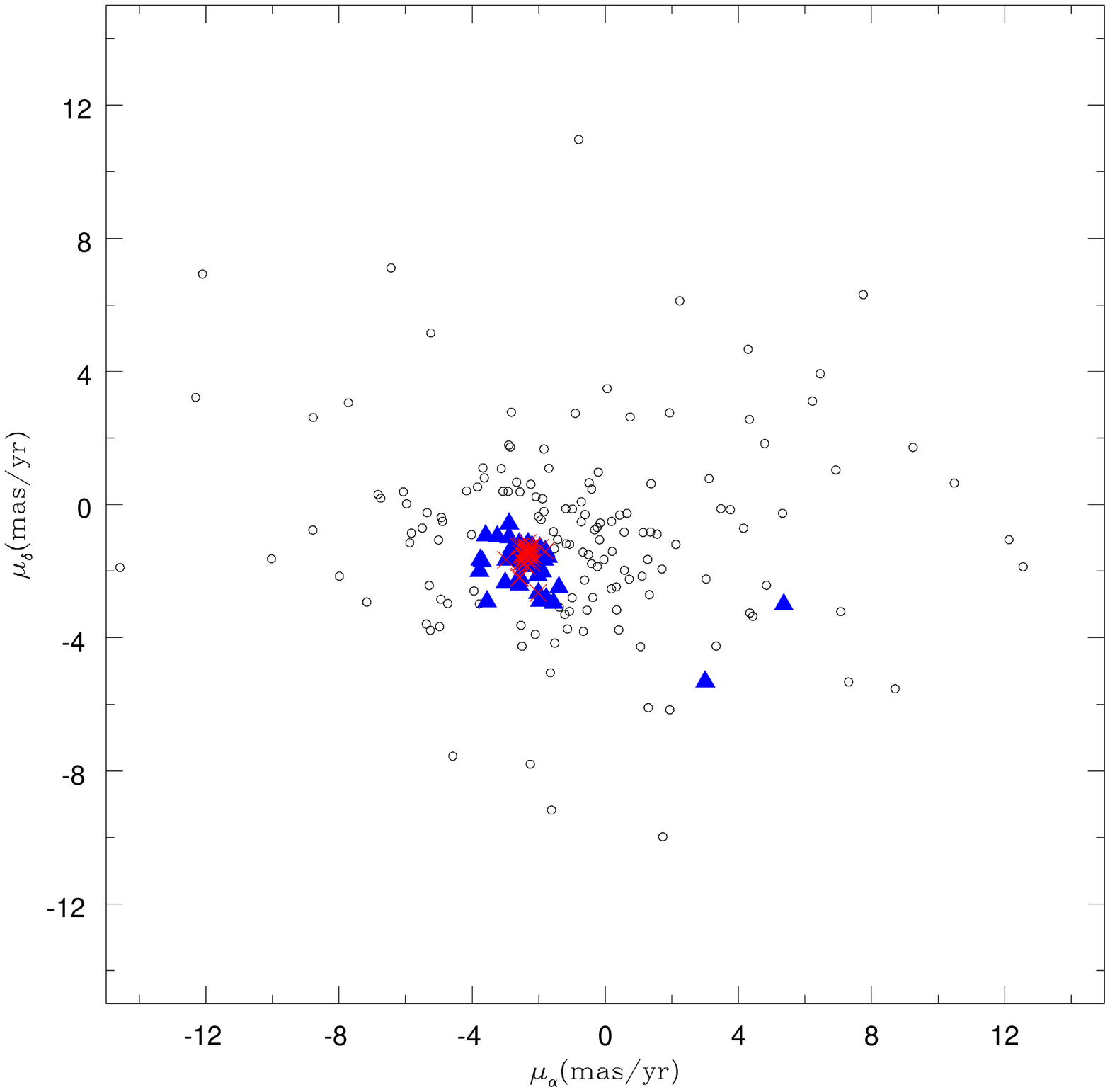}
\caption{The VPD of the proper motions for stars in the NGC 281 region. Open circles represent proper motions of 223 stars (cf. section 3.4) while triangles represent variables with membership probabilities $\ge$50\%. The crosses show  
present cluster members (both MS and PMS) with membership probabilities $\ge$50\%.
}
\end{figure}
\begin{figure}
\includegraphics[width=9cm]{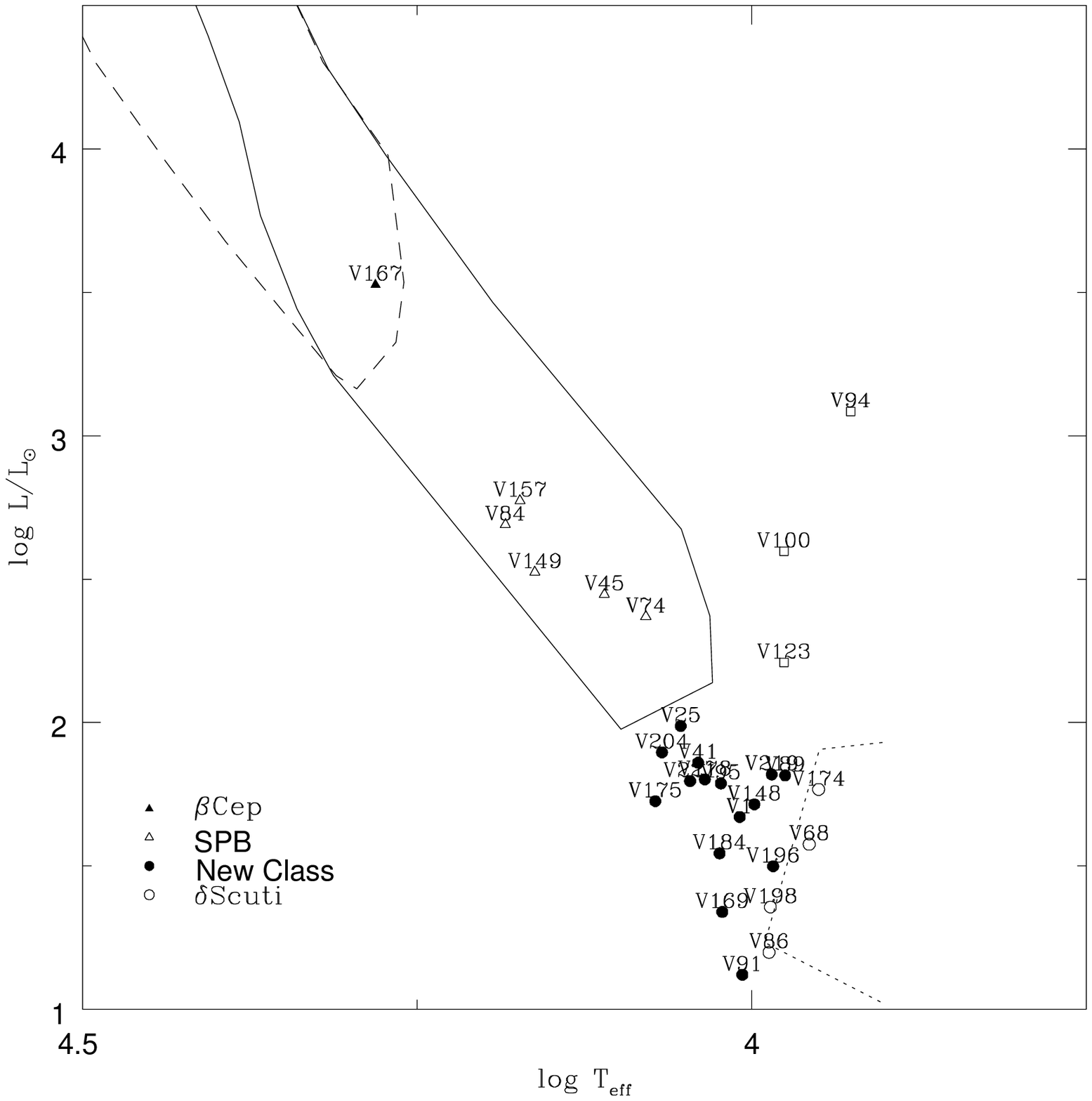}
\caption{ $\log(L/L_{\odot})/ \log T_{\rm eff}$ diagram for the probable MS variable stars identified in the present study.
The continuous curve represents the instability strip of SPB stars where as dotted curves shows the instability region of $\delta$ Scuti stars. The dashed curve shows the location of $\beta$ Cep stars (cf. Balona et al. 2011).
}
\end{figure}
\subsection{Kinematic data}
The proper motion values of stars have been taken from Gaia astrometric mission (Gaia Collaboration et al. 2018). For this we have taken data of stars which lie within
radial distance of 16 arcmin from center of the cluster NGC 281. The center of the cluster is assumed at RA=00h 52m 49s and Dec=56d 37m 48s (Sharma et al. 2012).
 For the membership probabilities of these stars,
we used the online Clusterix membership (http://clusterix.cerit-sc.cz/) estimation tool which uses the Non-Parametric method for the cluster-field separation in 2-dimensional proper motion space. This approach uses the kernel
estimation technique using a circular Gaussian kernel function to derive the data distributions as described
by Cabrera-Cano \& Alfaro (1990). The fundamental idea underlying this approach is the empirical determination
of the cluster and field star distributions without any assumption regarding their shape. The non-parametric method
works best for the open clusters where it is somewhat difficult to locate the center of the field region.
The proper motion cut-off was set to 15 mas/yr and radius of the cluster was considered as 8.0 arcmin. For field
region we have considered a region lying between 8 and 16 arcmin. 
Clusterix gave membership probabilities of 3748 stars.
The cross matched between present data and data (proper motions and membership probabilities of stars) given by Clusterix
reveals that 223 identified variables have membership probability.
The membership probability of variable stars along with proper motion values is given in Table 2.
Fig. 11 shows the vector point diagram (VPD) of the proper motions for stars of the NGC 281.
The highest concentrated area in the VPD gives 44 of 81 members identified from TCDs and CMD, and these 44 stars could be the most probable members of cluster. 
The 44 members distributed in the highest concentrated area in Fig. 11 have membership probabilities $\ge$50\%.
Out of 81 members, 37 stars with membership probabilities $<$ 50\% could be considered as the least probable members of the cluster
because of their locations in $V/V-I$ CMD and $J-H/H-K$ TCD.
There are 35 field stars with membership probabilities $\ge$50\%. Though the membership probabilities of all these 35 stars are $\ge$50\% and 33 of which are located where PMS stars are found to be lying in $V/V-I$ CMD, they could not be considered as members of the cluster because of their locations in $J-H/H-K$ TCD. 

%**********************
\begin{table*}
%\centering
\caption{The $VI$ data, amplitude and period of variables in the open cluster NGC 281. The Proper motion data were taken from Gaia (Gaia Collaboration et al. 2018).
The column Prob. refers to the membership probabilities of variable stars calculated using proper motion data.
}
\tiny
\begin{tabular}{llllccclllll}
\hline
ID  &  RA(2000)&   Dec(2000)&      $V$&   $I$&  $\mu_{RA}$  &$\mu_{Dec}$ & Prob.& Amplitude & Period& Class. \\
    &          &            &      (mag)&    (mag) & mas/yr & mas/yr& & (mag) &(days)&        \\
\hline
  V1   &  13.394370  & 56.588220 &  13.961$\pm$0.018 &  12.944$\pm$0.021 & -5.497  &  -0.706 &    0.22     &   0.032   &0.297                                        & MS    \\        
  V2   &  13.340740  & 56.511100 &  15.530$\pm$0.024 &  14.349$\pm$0.013 &  4.332  &   2.558 &      0.     &   0.045   &0.375, 0.407                                 & Field \\
  V3   &  13.418240  & 56.627450 &  15.534$\pm$0.020 &  14.650$\pm$0.021 & -5.376  &  -3.587 &    0.31     &   0.037   &0.222,0.212                                  & Field \\
  V4   &  13.362610  & 56.548570 &  18.254$\pm$0.051 &  16.958$\pm$0.018 &  -0.24  &  -1.861 &    0.23     &   0.076   &0.331,0.327                                  & Field \\
  V5   &  13.349920  & 56.541820 &  16.535$\pm$0.026 &  15.442$\pm$0.008 &  -4.89  &  -0.504 &    0.28     &   0.045   &0.495                                        & Field \\
  V6   &  13.455430  & 56.695880 &  15.845$\pm$0.023 &  14.694$\pm$0.009 & -0.207  &   0.974 &    0.08     &   0.103   &0.157                                        & Field \\
  V7   &  13.405920  & 56.625330 &  16.091$\pm$0.018 &  15.137$\pm$0.011 & -1.135  &  -3.733 &    0.34     &   0.035   &0.572                                        & Field \\
  V8   &  13.408430  & 56.634080 &  16.162$\pm$0.018 &  15.040$\pm$0.011 & -2.889  &  -0.987 &    0.62     &   0.045   &0.573                                        & Field \\
  V9   &  13.413880  & 56.642560 &  17.391$\pm$0.037 &   -               & -1.556  &  -0.809 &    0.36     &   0.052   &0.351                                        & Field \\
  V10  &  13.400810  & 56.624510 &  16.165$\pm$0.020 &  15.073$\pm$0.014 &  1.293  &  -6.098 &      0.     &   0.040   &0.583,0.441                                  & Field \\
  V11  &  13.385840  & 56.604010 &  15.376$\pm$0.021 &  14.236$\pm$0.017 &   1.73  &  -9.976 &      0.     &   0.043   &0.754,0.412                                  & Field \\
  V12  &  13.349280  & 56.555970 &  17.728$\pm$0.029 &  15.912$\pm$0.398 &  7.309  &  -5.325 &      0.     &   0.072   &0.811                                        & PMS   \\
  V13  &  13.317420  & 56.511180 &  12.224$\pm$0.019 &    -              & -1.888  &  -2.018 &    0.63     &   0.044   &0.579,0.634,0.614                            & Field \\
  V14  &  13.423000  & 56.666444 &  14.371$\pm$0.008 &  13.707$\pm$0.010 &  0.576  &  -1.972 &    0.17     &   0.046   &0.488,0.522                                  & Field \\
  V15  &  13.419056  & 56.661000 &  18.807$\pm$0.242 &  18.088$\pm$0.065 &  3.759  &  -0.154 &      0.     &   0.261   &0.129                                         & Field \\
  V16  &  13.397806  & 56.630806 &  15.236$\pm$0.008 &  14.425$\pm$0.008 & -1.985  &  -1.454 &    0.61     &   0.039   &0.492,0.499,0.441,0.434                       & Field\\
  V17  & 13.375028  & 56.598694 &  15.991$\pm$0.010 &   15.223$\pm$0.012 &  2.121  &  -1.196 &      0.     &   0.052  &0.628                                         & Field\\
  V18  & 13.388667  & 56.620694 &  15.668$\pm$0.011 &   14.215$\pm$0.009 & -2.199  &  -1.252 &    0.61     &   0.068  &0.751                                         & PMS  \\
  V19  & 13.334667  & 56.553167 &  16.538$\pm$0.041 &   15.352$\pm$0.010 & -1.781  &  -2.825 &    0.56     &   0.037  &0.448                                         & Field\\
  V20  & 13.401583  & 56.653056 &  14.328$\pm$0.006 &   13.774$\pm$0.008 & -0.659  &  -3.806 &    0.28     &   0.038  &0.454,0.427                                   & Field\\
  V21  & 13.413111  & 56.674639 &  17.765$\pm$0.032 &   16.409$\pm$0.014 &  0.749  &   2.632 &      0.     &   0.075  &0.334,0.331                                   & PMS  \\
  V22  & 13.321028  & 56.541028 &  16.825$\pm$0.025 &   15.807$\pm$0.013 &  0.191  &  -0.505 &    0.16     &   0.049  &0.487,0.530                                   & Field\\
  V23  & 13.315000  & 56.536750 &  14.161$\pm$0.006 &   13.325$\pm$0.009 & -5.242  &   5.157 &      0.     &   0.020  &0.530                                         & Field\\
  V24  & 13.394694  & 56.653278 &  17.153$\pm$0.021 &   16.062$\pm$0.011 & -0.409  &   0.468 &    0.12     &   0.069  &0.447                                         & Field\\
  V25  & 13.329500  & 56.559972 &  13.526$\pm$0.005 &   13.103$\pm$0.006 & -2.191  &  -1.506 &    0.64     &   0.028  &0.468                                         & MS   \\
  V26  & 13.322417  & 56.552472 &  15.908$\pm$0.013 &   14.796$\pm$0.009 &  1.105  &  -2.151 &     0.1     &   0.035  &0.750,0.816                                   & Field\\
  V27  & 13.332667  & 56.567500 &  14.648$\pm$0.007 &   13.990$\pm$0.007 & -2.273  &  -1.534 &    0.65     &   0.037  &0.308                                         & Field\\
  V28  & 13.362556  & 56.616000 &  15.407$\pm$0.010 &   14.550$\pm$0.012 &  1.378  &   0.626 &      0.     &   0.059  &0.527                                         & Field\\
  V29  & 13.388639  & 56.654694 &  17.928$\pm$0.039 &   16.342$\pm$0.031 &  -      &   -     &      -      &   0.160  &0.485,0.527                                   & PMS  \\
  V30  & 13.324583  & 56.565278 &  17.918$\pm$0.066 &   16.386$\pm$0.013 & -2.016  &  -2.658 &    0.61     &   0.078  &0.306                                         & PMS  \\
  V31  & 13.347389  & 56.599528 &  15.858$\pm$0.023 &   14.865$\pm$0.008 &  9.249  &   1.719 &      0.     &   0.059  &0.481,901                                     & Field\\
  V32  & 13.410944  & 56.693083 &  18.270$\pm$0.045 &   17.206$\pm$0.025 & -2.821  &   2.777 &      0.     &   0.127  &0.481                                         & Field\\
  V33  & 13.390361  &  56.66475 &  16.673$\pm$0.017 &   15.657$\pm$0.012 &  0.184  &  -2.532 &    0.21     &   0.051  &0.497,0.482                                   & Field\\
  V34  & 13.336667  & 56.587139 &  13.894$\pm$0.009 &   12.986$\pm$0.006 &  3.333  &  -4.249 &    0.45     &   0.058  &0.654                                         & Field\\
  V35  & 13.323556  & 56.568694 &  16.602$\pm$0.017 &   14.775$\pm$0.011 & -3.674  &   1.102 &    0.22     &   0.044  &0.449                                         & PMS  \\
  V36  & 13.375361  & 56.644583 &  17.438$\pm$0.062 &   16.596$\pm$0.032 & -4.941  &  -2.843 &    0.42     &   0.201  &0.788,0.447,0.737                             & Field\\
  V37  & 13.299611  & 56.537639 &  14.727$\pm$0.007 &   13.745$\pm$0.008 &  -      &  -      &      -       &   0.035  &0.354,0.386                                   & Field\\
  V38  & 13.354639  & 56.618389 &  17.470$\pm$0.033 &   15.956$\pm$0.012 & -2.243  &  -1.486 &    0.65     &   0.087  &0.722,0.779                                   & PMS  \\
  V39  & 13.324583  & 56.580528 &  17.144$\pm$0.034 &   15.457$\pm$0.008 & -1.381  &  -3.081 &    0.45     &   0.058  &0.429,0.732,0.418                             & PMS  \\
  V40  & 13.347361  & 56.617222 &  15.168$\pm$0.017 &   14.335$\pm$0.007 & -5.007  &  -1.064 &    0.28     &   0.076  &0.489                                         & Field\\
  V41  & 13.289889  & 56.535167 &  13.682$\pm$0.008 &   12.922$\pm$0.006 &   0.33  &  -2.475 &    0.19     &   0.042  &0.572                                         & MS   \\
  V42  & 13.326389  & 56.590611 &  16.111$\pm$0.029 &   15.167$\pm$0.013 & -5.968  &   0.027 &    0.28     &   0.066  &0.735                                         & Field\\
  V43  & 13.385611  & 56.679278 &  17.504$\pm$0.051 &   16.099$\pm$0.017 & -1.615  &  -9.175 &      0.     &   0.041  &0.321                                         & Field\\
  V44  & 13.356528  & 56.638389 &  14.966$\pm$0.007 &   13.567$\pm$0.006 & -1.843  &  -0.211 &    0.28     &   0.054  &0.441,0.504                                   & MS   \\
  V45  & 13.385694  & 56.683417 &  12.448$\pm$0.008 &   12.387$\pm$0.012 & -2.225  &  -1.607 &    0.66     &   0.053  &0.467                                         & Field\\
  V46  & 13.336500  & 56.612444 &  13.829$\pm$0.013 &   12.921$\pm$0.009 &  0.057  &   3.486 &      0.     &   0.082  &0.489                                         & Field\\
  V47  & 13.278389  & 56.529556 &  13.746$\pm$0.012 &   12.368$\pm$0.009 & -3.757  &  -1.657 &    0.56     &   0.049  &0.347                                         & Field\\
  V48  & 13.275667  & 56.527500 &  17.048$\pm$0.044 &   15.738$\pm$0.011 &  0.207  &  -1.404 &    0.18     &   0.069  &0.482,0.492                                   & Field\\
  V49  & 13.388472  & 56.694111 &  16.072$\pm$0.011 &   15.083$\pm$0.009 & -8.785  &  -0.763 &      0.     &   0.068  &0.475,0.507                                   & PMS  \\
  V50  & 13.329250  & 56.608333 &  14.804$\pm$0.008 &   13.892$\pm$0.006 &  2.239  &   6.123 &      0.     &   0.035  &0.412,0.423,0.318,0.245                & Field\\
  V51  & 13.346583  & 56.635944 &  16.205$\pm$0.013 &   15.203$\pm$0.008 &   1.93  &   2.755 &      0.     &   0.067  &1.035                                         & Field\\
  V52  & 13.375139  & 56.677611 &  18.425$\pm$0.060 &   17.280$\pm$0.049 & -2.084  &   0.236 &    0.21     &   0.105  &0.456                                         & Field\\
  V53  & 13.276667  & 56.535389 &  18.757$\pm$0.074 &   16.933$\pm$0.023 & -2.579  &  -1.517 &    0.68     &   0.140  &0.468                                         & Field\\
  V54  & 13.323639  & 56.605444 &  17.450$\pm$0.092 &   16.260$\pm$0.011 & -3.697  &   -1.71 &    0.57     &   0.105  &0.476,0.864,0.516                             & Field\\
  V55  & 13.367833  & 56.672722 &  18.308$\pm$0.053 &   16.601$\pm$0.013 & -2.273  &  -1.315 &    0.62     &   0.296  &0.511,1.02                                    & PMS  \\
  V56  & 13.321417  & 56.610500 &  17.201$\pm$0.020 &   15.708$\pm$0.013 & -2.175  &  -1.382 &    0.62     &   0.104  &0.719,0.475                                   & Field\\
  V57  & 13.371333  & 56.686944 &  17.023$\pm$0.027 &   15.351$\pm$0.029 & -1.723  &  -1.578 &    0.57     &   0.058  &0.601,0.648                                   & Field  \\
  V58  & 13.369972  & 56.685972 &  15.419$\pm$0.009 &   13.792$\pm$0.009 & -0.149  &  -0.552 &    0.15     &   0.042  &0.482,0.497,0.533                            & PMS  \\
  V59  & 13.297333  & 56.580694 &  17.802$\pm$0.037 &   16.491$\pm$0.012 & -2.581  &  -1.163 &    0.64     &   0.069  &0.423,0.449,0.433,0.482                      & Field\\
  V60  & 13.298278  & 56.583806 &  16.130$\pm$0.017 &   15.044$\pm$0.006 & 12.129  &  -1.055 &      0.     &   0.056  &0.468,0.684                                  & Field\\
  V61  & 13.362944  & 56.679222 &  15.964$\pm$0.014 &   15.119$\pm$0.016 & -1.076  &   -3.21 &     0.4     &   0.062  &0.828,0.451                                  & Field\\
  V62  & 13.280194  & 56.561361 &  13.715$\pm$0.014 &   12.917$\pm$0.008 & -3.004  &  -2.357 &    0.66     &   0.017  &0.325,0.251                                  & Field\\
  V63  & 13.282667  & 56.565722 &  17.248$\pm$0.024 &   16.248$\pm$0.016 & -2.503  &  -4.255 &    0.34     &   0.069  &0.499,0.532                                  & Field\\
  V64  & 13.285611  & 56.572167 &  15.434$\pm$0.011 &   14.224$\pm$0.006 & -2.231  &  -1.469 &    0.64     &   0.061  &0.336                                        & Field\\
  V65  & 13.288722  & 56.580389 &  16.583$\pm$0.024 &   15.651$\pm$0.009 & -1.841  &   1.671 &     0.2     &   0.056  &0.432,0.336,0.429,0.499                      & Field\\
  V66  & 13.301639  & 56.599722 &  17.546$\pm$0.037 &   16.372$\pm$0.023 &  -2.03  &  -2.126 &    0.64     &   0.097  &0.516,0.864,0.489,0.828                      & Field\\
  V67  & 13.358639  & 56.683389 &  16.750$\pm$0.018 &   15.860$\pm$0.011 & -0.986  &  -0.131 &    0.15     &   0.044  &0.828,0.935,0.481,1.22                       & Field\\
  V68  & 13.271528  & 56.556500 &  13.949$\pm$0.007 &   13.584$\pm$0.006 &  0.345  &  -3.165 &    0.23     &   0.043  &0.217,0.336                                  & MS   \\
  V69  & 13.264167  & 56.547889 &  14.910$\pm$0.007 &   14.127$\pm$0.007 &  0.572  &  -0.828 &    0.17     &   0.023  &0.271                                        & Field\\
  V70  & 13.361861  & 56.690722 &  17.783$\pm$ 0.04 &   16.716$\pm$0.017 & -0.411  &  -1.775 &    0.25     &   0.104  &0.533,0.919                                  & Field\\
  V71  & 13.284639  & 56.582861 &  15.994$\pm$0.010 &   15.173$\pm$0.012 & -3.836  &   0.531 &    0.28     &   0.047  &0.467,0.794                                  & Field\\
  V72  & 13.361444  & 56.704306 &  15.658$\pm$0.009 &   14.968$\pm$0.009 & -0.673  &  -1.428 &    0.27     &   0.042  &0.459,0.481                                  & Field\\
  V73  & 13.276639  & 56.581278 &  15.198$\pm$0.007 &   14.346$\pm$0.006 &  1.137  &  -0.837 &    0.14     &   0.058  &0.469                                        & Field\\
  V74  & 13.318472  & 56.642528 &  12.717$\pm$0.008 &   12.369$\pm$0.007 & -2.254  &   -1.44 &    0.64     &   0.011  &0.273,0.239,0.233                             & MS   \\
  V75  & 13.346361  & 56.683611 &  16.608$\pm$0.023 &   15.530$\pm$0.010 &  4.791  &   1.832 &      0.     &   0.073  &0.171,0.8302,0.498,0.460                     & Field\\
  V76  & 13.236028  & 56.526250 &  17.417$\pm$0.025 &   16.308$\pm$0.013 & -2.855  &    1.73 &    0.22     &   0.129  &0.769                                        & Field  \\
  V77  & 13.330972  & 56.665333 &  14.478$\pm$0.012 &   13.684$\pm$0.009 & -1.839  &  -1.668 &     0.6     &   0.013  &0.308,0.318                                  & PMS\\
  V78  & 13.359389  & 56.709222 &  14.062$\pm$0.008 &   12.353$\pm$0.011 & -2.237  &    0.61 &    0.17     &   0.035  &0.314,0.710,0.332,0.512                      & PMS  \\
  V79  & 13.269944  & 56.580111 &  17.225$\pm$0.040 &   16.297$\pm$0.017 & -2.532  &  -3.626 &    0.44     &   0.055  &0.459,0.471                                  & Field\\
  V80  & 13.254417  & 56.560833 &  13.464$\pm$0.007 &   12.676$\pm$0.011 & -2.663  &   0.669 &    0.22     &   0.019  &0.325                                        & Field\\
  V81  & 13.280000  & 56.604917 &  15.397$\pm$0.021 &   14.473$\pm$0.007 & -1.558  &  -2.961 &     0.5     &   0.037  &0.338                                        & Field\\
  V82  & 13.304917  & 56.643000 &  15.239$\pm$0.007 &   14.307$\pm$0.006 & -1.515  &  -4.161 &    0.31     &   0.017  &0.249,0.332                                  & Field\\
  V83  & 13.351472  & 56.712806 &  17.087$\pm$0.024 &   16.205$\pm$0.013 & -3.783  &  -2.982 &    0.44     &   0.075  &0.475                                        & Field\\
  V84  & 13.302139  & 56.643861 &  12.534$\pm$0.005 &   12.208$\pm$0.006 & -2.579  &  -2.209 &    0.68     &   0.072  &0.434                                        & MS   \\
  V85  & 13.276806  & 56.608778 &  16.946$\pm$0.020 &   15.738$\pm$0.011 & -3.248  &  -0.959 &     0.6     &   0.060  &0.471,0.523                                  & Field\\
  V86  & 13.278083  & 56.612972 &  15.255$\pm$0.008 &   14.630$\pm$0.008 & -2.451  &  -1.532 &    0.67     &   0.021  &0.234,0.305,0.302                                  & MS   \\
  V87  & 13.314944  & 56.671000 &  17.611$\pm$0.026 &   16.157$\pm$0.019 &  -0.72  &  -0.515 &    0.16     &   0.088  &0.560                                        & PMS  \\
  V88  & 13.336028  & 56.702222 &  18.896$\pm$0.127 &   17.407$\pm$0.031 & -       &  -      &      -       &   0.238  &0.332                                        & Field\\
  V89  & 13.236778  & 56.559639 &  13.344$\pm$0.007 &   12.876$\pm$0.007 & -1.069  &  -1.191 &    0.32     &   0.017  &0.328,0.332,0.343,0.498,0.348,7.745          & MS   \\
  V90  & 13.323944  & 56.687389 &  17.106$\pm$0.023 &   16.158$\pm$0.012 & -3.127  &   1.085 &    0.22     &   0.054  &0.446                                        & Field\\
  V91  & 13.255417  & 56.588111 &  15.329$\pm$0.009 &   14.830$\pm$0.007 & -2.382  &   -1.45 &    0.66     &   0.058  &0.525                                        & MS   \\
  V92  & 13.252111  & 56.584556 &  16.109$\pm$0.011 &   14.992$\pm$0.006 &  3.477  &  -0.124 &      0.     &   0.065  &0.429,0.435                                  & Field\\
  V93  & 13.330389  & 56.698889 &  16.921$\pm$0.018 &   15.370$\pm$0.013 &  7.755  &   6.308 &      0.     &   0.076  &0.560                                       & PMS   \\
  V94  & 13.339917  & 56.714556 &  10.036$\pm$0.007 &    9.662$\pm$0.034 & 12.553  &  -1.867 &      0.     &   0.066  &0.524                                        & MS   \\
  V95  & 13.295972  & 56.650639 &  13.924$\pm$0.006 &   13.255$\pm$0.005 & -2.292  &  -1.357 &    0.64     &   0.022  &0.321,0.332,0.339                            & MS   \\
  V96  & 13.275417  & 56.621278 &  16.927$\pm$0.039 &   15.382$\pm$0.009 &  -2.98  &  -1.661 &    0.68     &   0.073  &1.253                                        & PMS  \\
  V97  & 13.238611  & 56.574194 &  17.309$\pm$0.023 &   16.365$\pm$0.017 & -3.592  &  -0.925 &    0.56     &   0.081  &4.225                                        & Field\\
  V98  & 13.280944  & 56.636194 &  13.708$\pm$0.007 &   12.240$\pm$0.006 & -6.747  &   0.192 &    0.42     &   0.029  &1.114,0.400,0.324                            & PMS  \\
  V99  & 13.307472  & 56.681222 &  16.263$\pm$0.028 &   15.258$\pm$0.008 & -1.647  &   -5.05 &      0.     &   0.046  &1.423,0.501,0.533                            & Field\\
  V100 & 13.218417  & 56.551361 &  11.400$\pm$0.006 &   11.202$\pm$0.006 & -2.102  &  -3.898 &    0.38     &   0.017  &0.251                                        & MS   \\
  V101 & 13.270972  & 56.628306 &  15.786$\pm$0.011 &   14.820$\pm$0.009 &  4.158  &  -0.711 &      0.     &   0.020  &0.376,0.243                                  & Field\\
  V102 & 13.230389  & 56.574083 &  17.661$\pm$0.034 &   16.249$\pm$0.013 & -2.577  &  -2.172 &    0.68     &   0.101  &0.487                                        & PMS  \\
  V103 & 13.268556  & 56.630250 &  17.370$\pm$0.025 &   15.702$\pm$0.015 & -8.779  &   2.618 &      0.     &   0.109  &0.823                                       & PMS   \\
  V104 & 13.269750  & 56.634861 &  15.717$\pm$0.009 &   14.678$\pm$0.007 &  3.006  &  -5.316 &    0.73     &   0.081  &0.454,1.283                                  & Field\\
  V105 & 13.263083  & 56.630806 &  15.122$\pm$0.008 &   14.588$\pm$0.007 & -2.395  &  -1.418 &    0.66     &   0.019  &0.235                                        & Field\\
  V106 & 13.274167  & 56.656639 &  15.564$\pm$0.008 &   14.520$\pm$0.006 & -2.585  &  -2.415 &    0.67     &   0.064  &0.530,1.28                                   & Field\\
  V107 & 13.280750  & 56.667028 &  14.961$\pm$0.008 &   13.804$\pm$0.007 & -1.395  &  -2.482 &    0.52     &   0.025  &0.247,0.315                                  & Field\\
  V108 & 13.317944  & 56.721778 &  15.895$\pm$0.012 &   14.500$\pm$0.009 & -6.435  &   7.109 &      0.     &   0.071  &0.789,0.731,2.167                             & PMS  \\
  V109 & 13.220611  & 56.580917 &  16.145$\pm$0.029 &   12.837$\pm$0.005 &  1.557  &  -0.889 &    0.06     &   0.065  &0.321                                       & PMS   \\
  V110 & 13.222778  & 56.586278 &  18.278$\pm$0.089 &   17.050$\pm$0.037 & -0.621  &   -2.27 &    0.34     &   0.090  &0.474,0.787                                 & Field \\
  V111 & 13.243694  & 56.619639 &  15.269$\pm$0.008 &   13.759$\pm$0.005 & -0.244  &  -0.687 &    0.16     &   0.021  &0.367,0.354,0.441                           & PMS   \\
\hline
\end{tabular}
\end{table*}
%***************************************************************************************************************************88
\setcounter{table}{1}
\begin{table*}
%\centering
\caption{Continued.
}
\tiny
\begin{tabular}{lllllccclllll}
\hline
ID  &  RA(2000)&   Dec(2000)&      $V$&   $I$&  $\mu_{RA}$  &$\mu_{Dec}$ & Prob.& Amplitude & Period& Class. \\
    &          &            &      (mag)&    (mag) & mas/yr & mas/yr& & (mag) &(days)&        \\
\hline

  V112    & 13.278083  &56.670528 &15.678$\pm$0.008 &  14.853$\pm$0.006 & -2.807  &  -1.294 &   0.66 &   0.030    &0.315                                       & Field \\
  V113    & 13.202000  &56.564306 &15.952$\pm$0.013 &  15.068$\pm$0.008 & -2.919  &   0.396 &   0.28 &   0.096    &14.399                                      & Field \\
  V114    & 13.226889  &56.601361 &18.373$\pm$0.050 &  16.878$\pm$0.029 & -2.486  &  -1.704 &   0.68 &   0.132    &0.572                                       & PMS   \\
  V115    & 13.276500  &56.675028 &15.895$\pm$0.037 &  15.106$\pm$0.007 & -3.946  &  -2.593 &   0.46 &   0.041    &0.561                                       & Field \\
  V116    & 13.296611  &56.704944 &16.011$\pm$0.012 &  15.012$\pm$0.008 & 10.488  &   0.648 &     0. &   0.064    &1.068,0.489,0.525,1.147                     & Field \\
  V117    & 13.307306  &56.721611 &15.929$\pm$0.019 &  14.887$\pm$0.010 & -3.074  &   0.398 &   0.28 &   0.067    &0.768                                       & Field \\
  V118    & 13.292306  &56.701083 &16.311$\pm$0.034 &  14.978$\pm$0.010 & -1.699  &   1.091 &   0.13 &   0.046    &0.524                                       & Field \\
  V119    & 13.181250  &56.540222 &15.303$\pm$0.021 &  14.566$\pm$0.007 & -3.777  &  -1.995 &   0.55 &   0.018    &0.211                                       & Field \\
  V120    & 13.237139  &56.623556 &17.417$\pm$0.045 &  16.015$\pm$0.018 &-14.581  &  -1.891 &     0. &   0.105    &0.541,3.363,1.0649                          & PMS   \\
  V121    & 13.201694  &56.571583 &17.715$\pm$0.059 &  16.265$\pm$0.014 & -7.989  &  -2.151 &     0. &   0.111    &0.747,1.248                                 & PMS   \\
  V122    & 13.283194  &56.691722 &15.982$\pm$0.009 &  15.021$\pm$0.007 &-10.029  &  -1.629 &     0. &   0.036    &0.455,1.286,0.533,1.177                     & Field \\
  V123    & 13.226056  &56.610611 &12.545$\pm$0.016 &  12.032$\pm$0.004 & -1.531  &  -1.328 &   0.49 &   0.049    &0.760                                       & MS    \\
  V124    & 13.287778  &56.700750 &18.525$\pm$0.110 &  16.635$\pm$0.047 & -1.797  &  -1.401 &   0.56 &   0.392    &4.0414                                      & PMS   \\
  V125    & 13.203333  &56.579972 &18.760$\pm$0.078 &  16.991$\pm$0.019 & -2.396  &  -1.546 &   0.67 &   0.188    &0.543                                       & PMS   \\
  V126    & 13.259917  &56.670000 &14.740$\pm$0.008 &  14.153$\pm$0.006 & -2.339  &  -1.386 &   0.64 &   0.015    &0.061,0.531                                  & PMS   \\
  V127    & 13.216833  &56.609694 &15.803$\pm$0.009 &  14.787$\pm$0.007 &  0.722  &  -2.241 &   0.16 &   0.045    &0.521,0.504                                & Field \\
  V128    & 13.240056  &56.644250 &15.557$\pm$0.006 &  14.143$\pm$0.005 & -2.343  &  -1.431 &   0.65 &   0.017    &0.251,0.253,0.249                          & PMS   \\
  V129    & 13.187167  &56.570806 &14.260$\pm$0.008 &  13.553$\pm$0.005 & -5.351  &  -0.243 &   0.25 &   0.012    &0.237                                      & Field \\
  V130    & 13.173333  &56.550944 &15.188$\pm$0.006 &  14.330$\pm$0.006 &-12.102  &   6.927 &     0. &   0.017    &0.290                                      & Field \\
  V131    & 13.203030  &56.595610 &18.480$\pm$0.074 &      -            & -3.555  &  -2.904 &    0.5 &   0.181    &0.453,0.883,0.447                          & Field \\
  V132    & 13.205056  &56.603056 &17.725$\pm$0.039 &  16.961$\pm$0.041 &  -2.01  &  -0.357 &   0.35 &   0.155    &0.516,0.533                                & Field \\
  V133    & 13.212694  &56.615806 &13.472$\pm$0.009 &  12.763$\pm$0.006 & -2.329  &  -1.424 &   0.65 &   0.040    &0.356,0.571                                & PMS   \\
  V134    & 13.252194  &56.676028 &17.145$\pm$0.020 &  16.058$\pm$0.017 &  4.846  &  -2.423 &   0.35 &   0.041    &0.322                                      & Field \\
  V135    & 13.214750  &56.624694 &14.859$\pm$0.017 &  14.203$\pm$0.022 & -2.317  &  -1.364 &   0.64 &   0.051    &0.541                                      & Field \\
  V136    & 13.248972  &56.674944 &16.654$\pm$0.027 &  15.891$\pm$0.012 & -4.976  &  -3.663 &   0.28 &   0.034    &0.433                                      & Field \\
  V137    & 13.232778  &56.656556 &15.072$\pm$0.007 &  14.218$\pm$0.006 & -2.326  &  -1.451 &   0.65 &   0.032    &0.479,0.545,1.326                          & PMS   \\
  V138    & 13.228500  &56.650861 &15.544$\pm$0.015 &  13.909$\pm$0.007 & -0.721  &   0.083 &   0.13 &   0.026    &0.615,0.686                                & PMS   \\
  V139    & 13.214556  &56.630917 &14.734$\pm$0.020 &  13.575$\pm$0.014 & -2.306  &  -1.538 &   0.66 &   0.021    &0.287                                      & Field \\
  V140    & 13.217222  &56.639472 &16.544$\pm$0.015 &  15.728$\pm$0.011 & -2.886  &  -0.572 &   0.53 &   0.040    &0.658, 0.631                               & Field \\
  V141    & 13.167694  &56.569250 &14.380$\pm$0.007 &  13.656$\pm$0.007 & -4.166  &   0.413 &   0.28 &   0.013    &0.339                                      & Field \\
  V142    & 13.242861  &56.679806 &14.690$\pm$0.009 &  13.884$\pm$0.008 & -5.255  &  -3.775 &   0.26 &   0.022    &0.427,0.485                                & Field \\
  V143    & 13.203639  &56.625417 & 9.769$\pm$0.025 &   9.612$\pm$0.014 & -2.366  &  -1.437 &   0.66 &   0.059    &0.575                                      & MS    \\
  V144    & 13.243528  &56.686000 &13.687$\pm$0.008 &  12.326$\pm$0.009 &  1.061  &   -4.27 &   0.31 &   0.022    &0.443,0.554,0.423                          & PMS   \\
  V145    & 13.207417  &56.635583 &13.609$\pm$0.036 &  13.327$\pm$0.009 & -2.347  &  -1.495 &   0.66 &   0.027    &0.530,1.22                                 & MS    \\
  V146    & 13.161417  &56.569083 &15.752$\pm$0.013 &  14.937$\pm$0.008 & -7.714  &    3.06 &     0. &   0.015    &0.364                                      & Field \\
  V147    & 13.155250  &56.560139 &17.427$\pm$0.028 &  15.868$\pm$0.015 &  4.293  &   4.668 &     0. &   0.137    &3.302,4.474                                & PMS   \\
  V148    & 13.185333  &56.607833 &13.904$\pm$0.014 &  13.532$\pm$0.006 & -2.432  &  -1.601 &   0.67 &   0.039    &0.469,0.332                                & MS    \\
  V149    & 13.201139  &56.631694 &12.618$\pm$0.014 &  12.286$\pm$0.012 &  -2.37  &  -1.479 &   0.66 &   0.023    &0.554,0.539                                & MS    \\
  V150    & 13.189917  &56.620972 &14.799$\pm$0.007 &  13.725$\pm$0.007 & -2.279  &  -1.494 &   0.65 &   0.029    &0.334,0.490,0.460                          & PMS   \\
  V151    & 13.201778  &56.640028 &15.887$\pm$0.011 &  14.636$\pm$0.007 & -2.404  &  -1.552 &   0.67 &   0.050    &1.095,1.23                                 & PMS   \\
  V152    & 13.184611  &56.618056 &11.293$\pm$0.010 &  10.988$\pm$0.004 & -2.371  &  -1.282 &   0.64 &   0.016    &0.460,0.342                                & PMS   \\
  V153    & 13.168778  &56.597500 &13.467$\pm$0.007 &  12.692$\pm$0.005 & -2.314  &   -1.58 &   0.66 &   0.018    &0.111                                      & Field \\
  V154    & 13.188833  &56.628194 &14.703$\pm$0.008 &  13.601$\pm$0.007 & -2.404  &  -1.446 &   0.66 &   0.060    &0.550,0.513,1.094                          & PMS   \\
  V155    & 13.202389  &56.648917 &16.273$\pm$0.015 &  15.507$\pm$0.014 & -5.289  &  -2.426 &   0.42 &   0.084    &0.710                                      & Field \\
  V156    & 13.189528  &56.631917 &13.943$\pm$0.011 &  13.450$\pm$0.007 &  -2.36  &  -1.417 &   0.66 &   0.041    &0.946,0.492                                & PMS   \\
  V157    & 13.180361  &56.620000 &12.137$\pm$0.012 &  11.905$\pm$0.005 & -2.484  &  -1.503 &   0.67 &   0.034    &0.540,0.520                                & MS    \\
  V158    & 13.191611  &56.639417 &15.208$\pm$0.021 &  13.733$\pm$0.006 & -4.019  &  -0.898 &   0.47 &   0.027    &0.526                                      & PMS   \\
  V159    & 13.152389  &56.583083 &15.358$\pm$0.016 &  14.599$\pm$0.007 & -6.066  &   0.384 &   0.28 &   0.083    &4.22                                       & Field \\
  V160    & 13.153639  &56.587778 &16.498$\pm$0.021 &  15.270$\pm$0.010 &  0.406  &  -3.766 &   0.26 &   0.163    &0.635,3.15,1.99,1.42                       & Field \\
  V161    & 13.191528  &56.648278 &14.613$\pm$0.006 &  13.785$\pm$0.006 &  8.711  &  -5.531 &     0. &   0.018    &0.520,0.538                                & Field \\
  V162    & 13.132417  &56.564167 &17.129$\pm$0.021 &  15.933$\pm$0.012 & -4.581  &   -7.56 &     0. &   0.127    &0.475                                      & Field \\
  V163    & 13.158000  &56.601944 &16.311$\pm$0.012 &  15.256$\pm$0.008 & -2.505  &  -1.543 &   0.67 &   0.065    &0.830                                      & Field \\
  V164    & 13.188667  &56.650472 &19.067$\pm$0.260 &  17.564$\pm$0.035 & -2.326  &  -1.157 &   0.61 &   0.182    &0.463                                      & PMS   \\
  V165    & 13.185139  &56.646028 &15.815$\pm$0.011 &  14.574$\pm$0.007 & -2.283  &  -1.374 &   0.64 &   0.105    &0.618,0.593,0.657,0.630                    & PMS   \\
  V166    & 13.192500  &56.658194 &14.881$\pm$0.008 &  13.625$\pm$0.005 &  0.653  &  -0.263 &   0.16 &   0.019    &0.682,0.651                                & PMS   \\
  V167    & 13.151000  &56.600222 &10.779$\pm$0.008 &  10.587$\pm$0.012 & -2.595  &  -1.254 &   0.65 &   0.023    &0.251,0.247                                & MS    \\
  V168    & 13.151556  &56.609667 &15.144$\pm$0.009 &  14.350$\pm$0.006 &  -2.43  &  -1.371 &   0.65 &   0.018    &0.1045                                     & Field \\
  V169    & 13.139250  &56.594083 &14.769$\pm$0.009 &  14.349$\pm$0.006 & -2.261  &  -1.328 &   0.63 &   0.015    &0.429,0.411,0.442,0.294                    & MS    \\
  V170    & 13.161556  &56.627167 &14.353$\pm$0.009 &  13.463$\pm$0.006 &  4.432  &  -3.352 &   0.37 &   0.032    &0.699                                      & Field \\
  V171    & 13.163139  &56.630444 &16.049$\pm$0.010 &  14.790$\pm$0.008 & -2.214  &  -1.744 &   0.67 &   0.098    &0.514,1.27                                 & PMS   \\
  V172    & 13.169556  &56.645500 &15.751$\pm$0.011 &  15.000$\pm$0.010 &  -1.96  &  -2.901 &   0.56 &   0.023    &0.639                                      & Field \\
  V173    & 13.111889  &56.561833 &17.021$\pm$0.028 &  15.916$\pm$0.009 &  7.079  &  -3.215 &     0. &   0.098    &0.459                                      & Field \\
  V174    & 13.139000  &56.604528 &13.425$\pm$0.012 &  12.924$\pm$0.006 & -2.223  &  -1.303 &   0.62 &   0.020    &0.235,0.207,0.423                                & MS    \\
  V175    & 13.220806  &56.727778 &14.337$\pm$0.011 &  13.651$\pm$0.009 & -0.317  &  -0.755 &   0.16 &   0.032    &0.397,0.469                                & MS    \\
  V176    & 13.161389  &56.642556 &15.668$\pm$0.010 &  14.739$\pm$0.006 & -5.872  &  -1.146 &   0.19 &   0.026    &0.625                                      & Field \\
  V177    & 13.182528  &56.674722 &17.048$\pm$0.023 &  16.169$\pm$0.013 &  1.277  &  -1.646 &   0.09 &   0.032    &0.545,0.403                                & Field \\
  V178    & 13.174222  &56.666472 &13.679$\pm$0.014 &  13.458$\pm$0.005 & -2.307  &  -1.299 &   0.63 &   0.026    &0.644,0.658,0.416                          & MS    \\
  V179    & 13.135278  &56.609444 &15.120$\pm$0.009 &  14.580$\pm$0.007 & -2.878  &  -1.569 &   0.68 &   0.056    &0.503,1.096                                & Field \\
  V180    & 13.213250  &56.729417 &17.177$\pm$0.024 &  16.217$\pm$0.015 &  0.435  &  -0.315 &   0.16 &   0.021    &0.481,0.435                                & Field \\
  V181    & 13.190917  &56.700778 &16.013$\pm$0.017 &  14.820$\pm$0.007 &  4.343  &  -3.254 &   0.34 &   0.054    &0.450                                      & Field \\
  V182    & 13.146306  &56.643361 &16.133$\pm$0.010 &  14.923$\pm$0.015 &   -2.9  &   1.791 &   0.22 &   0.033    &0.435,0.751                                & Field \\
  V183    & 13.130361  &56.620167 &17.300$\pm$0.052 &  16.214$\pm$0.012 & -2.254  &  -7.795 &     0. &   0.055    &0.660                                      & Field \\
  V184    & 13.144639  &56.659694 &14.354$\pm$0.008 &  14.070$\pm$0.006 & -2.384  &  -1.322 &   0.65 &   0.027    &0.752                                      & MS    \\
  V185    & 13.112861  &56.615694 &17.578$\pm$0.027 &  16.465$\pm$0.019 & -0.039  &  -1.647 &    0.2 &   0.092    &0.474                                      & PMS   \\
  V186    & 13.125889  &56.635694 &15.591$\pm$0.011 &  14.844$\pm$0.007 & -0.503  &  -1.505 &   0.24 &   0.030    &0.447                                      & Field \\
  V187    & 13.193556  &56.738056 &14.492$\pm$0.008 &  13.114$\pm$0.010 & -2.364  &  -1.663 &   0.68 &   0.048    &0.510,0.477                                & Field \\
  V188    & 13.094694  &56.597611 &16.724$\pm$0.016 &  15.090$\pm$0.008 & -1.962  &  -1.299 &   0.57 &   0.125    &3.591                                      & PMS   \\
  V189    & 13.120167  &56.636556 &17.298$\pm$0.048 &  16.315$\pm$0.025 &  1.362  &  -0.823 &    0.1 &   0.062    &0.606                                      & Field \\
  V190    & 13.167889  &56.707472 &18.349$\pm$0.070 &  17.424$\pm$0.034 & -2.563  &   0.383 &   0.24 &   0.117    &0.515                                      & Field \\
  V191    & 13.069278  &56.707472 &15.046$\pm$0.009 &  14.159$\pm$0.009 &  1.704  &  -1.934 &     0. &   0.064    &0.456                                      & Field \\
  V192    & 13.168028  &56.713917 &17.189$\pm$0.026 &  16.395$\pm$0.016 & -4.925  &  -0.384 &   0.28 &   0.058    &0.499.0.245                                & Field \\
  V193    & 13.116028  &56.639722 &15.598$\pm$0.010 &  14.950$\pm$0.009 & -2.385  &  -1.477 &   0.66 &   0.044    &0.471,0.447,0.491,0.454                    & Field \\
  V194    & 13.086056  &56.597417 &16.107$\pm$0.026 &  15.176$\pm$0.023 &  1.325  &   -2.71 &     0. &   0.059    &0.539,0.530                                & Field \\
  V195    & 13.175667  &56.730944 &13.483$\pm$0.006 &  12.703$\pm$0.010 &  5.368  &  -2.997 &    0.5 &   0.034    &0.976                                      & Field \\
  V196    & 13.141917  &56.687722 &14.331$\pm$0.007 &  14.069$\pm$0.007 & -2.527  &  -1.242 &   0.65 &   0.036    &0.487                                      & MS    \\
  V197    & 13.091861  &56.614639 &16.785$\pm$0.030 &  15.965$\pm$0.012 & -6.828  &   0.302 &   0.43 &   0.047    &0.454                                      & Field \\
  V198    & 13.108000  &56.641861 &14.624$\pm$0.009 &  14.270$\pm$0.007 & -2.436  &  -1.789 &   0.68 &   0.047    &0.255,0.318,0.340                          & MS    \\
  V199    & 13.152778  &56.711944 &18.092$\pm$0.115 &  17.060$\pm$0.023 & -3.634  &   0.803 &   0.26 &   0.071    &0.536,0.488                                & Field \\
  V200    & 13.132500  &56.688556 &15.192$\pm$0.010 &  13.885$\pm$0.007 & -1.174  &  -1.174 &   0.37 &   0.037    &0.492,0.460,0.530                          & PMS   \\
  V201    & 13.159194  &56.734167 &16.063$\pm$0.012 &  15.240$\pm$0.011 &   1.94  &  -6.164 &     0. &   0.050    &0.527,0.758,0.478                          & Field \\
  V202    & 13.163278  &56.740722 &16.229$\pm$0.042 &  15.504$\pm$0.010 & -4.733  &   -2.97 &    0.4 &   0.056    &0.822                                      & Field \\
  V203    & 13.135611  &56.702917 &15.396$\pm$0.012 &  14.586$\pm$0.008 &   -     &     -   &    -   &   0.040    &0.486                                      & Field \\
  V204    & 13.103556  &56.656333 &13.720$\pm$0.006 &  13.298$\pm$0.006 & -1.428  &  -1.047 &    0.4 &   0.045    &0.446,0.460,0.768                          & MS    \\
  V205    & 13.099694  &56.651806 &17.618$\pm$0.029 &  16.557$\pm$0.026 & -0.374  &   -2.79 &   0.31 &   0.075    &0.544,0.515                                & Field \\
  V206    & 13.084639  &56.632722 &15.661$\pm$0.011 &  14.907$\pm$0.009 & -1.212  &  -3.296 &   0.41 &   0.050    &0.51,0.833,0.506,0.475                     & Field \\
  V207    & 13.051778  &56.659778 &16.696$\pm$0.024 &  15.703$\pm$0.013 &  6.461  &   3.933 &     0. &   0.180    &0.532,3.732,1.068                          & Field \\
  V208    & 13.057028  &56.667472 &16.277$\pm$0.013 &  15.181$\pm$0.008 &  6.927  &   1.042 &     0. &   0.096    &0.545,5.89,1.303                           & Field \\
  V209    & 13.111417  &56.674722 &16.853$\pm$0.021 &  15.623$\pm$0.021 & -0.609  &  -0.301 &   0.15 &   0.049    &0.627                                      & PMS   \\
  V210    & 13.128028  &56.699167 &18.299$\pm$0.150 &  17.220$\pm$0.030 &  -1.93  &  -0.453 &   0.38 &   0.089    &0.522,0.466,0.505                          & Field \\
  V211    & 13.153778  &56.740778 &16.728$\pm$0.017 &  15.374$\pm$0.013 &  3.121  &   0.781 &     0. &   0.104    &0.469,0.779                                & PMS   \\
  V212    & 13.118611  &56.689972 &16.775$\pm$0.019 &  15.677$\pm$0.010 &  6.225  &    3.11 &     0. &   0.053    &0.520                                      & Field \\
  V213    & 13.070833  &56.682556 &17.417$\pm$0.047 &  16.446$\pm$0.021 & -0.485  &   0.658 &   0.11 &   0.084    &0.442                                      & Field \\
  V214    & 13.132028  &56.713806 &15.873$\pm$0.009 &  14.971$\pm$0.008 & -0.799  &  10.971 &     0. &   0.032    &0.451,0.432                                & Field \\
  V215    & 13.085056  &56.649472 &17.134$\pm$0.066 &  15.897$\pm$0.012 &   -     &    -    &   -    &   0.077    &0.774,0.497,0.533,0.523                      & Field \\
  V216    & 13.118667  &56.699944 &18.797$\pm$0.244 &  17.651$\pm$0.051 &  -0.17  &  -1.057 &   0.17 &   0.119    &0.472,0.487,0.514                            & Field \\
  V217    & 13.052306  &56.602722 &16.869$\pm$0.022 &  15.848$\pm$0.013 &  5.328  &   -0.26 &     0. &   0.042    &0.412,0.443,0.491,0.721                      & Field \\
  V218    & 13.117750  &56.705917 &14.979$\pm$0.011 &  13.498$\pm$0.008 &  -2.36  &  -1.848 &   0.68 &   0.034    &0.559,0.467                                  & PMS   \\
  V219    & 13.057194  &56.622222 &13.570$\pm$0.009 &  12.951$\pm$0.007 &  3.031  &  -2.236 &     0. &   0.059    &0.508,0.858,1.038,0.460,0.475                & MS    \\
  V220    & 13.076722  &56.653250 &15.466$\pm$0.009 &  12.395$\pm$0.006 & -1.184  &  -0.125 &   0.17 &   0.036    &0.335,7.089                                  & PMS   \\
  V221    & 13.044167  &56.610194 &13.857$\pm$0.008 &  13.476$\pm$0.007 & -2.506  &  -1.485 &   0.67 &   0.053    &0.883                                        & MS    \\
  V222    & 13.066583  &56.643306 &16.538$\pm$0.018 &  15.562$\pm$0.013 & -1.884  &   0.176 &   0.21 &   0.054    &0.460,0.751                                  & Field \\
  V223    & 13.105806  &56.701389 &16.690$\pm$0.022 &  15.717$\pm$0.010 & -0.996  &    -2.8 &   0.42 &   0.070    &0.489,0.498,0.460,0.858                      & Field \\
  V224    & 13.091139  &56.680139 &15.576$\pm$0.010 &  14.703$\pm$0.009 & -0.546  &  -3.169 &   0.33 &   0.052    &0.832,0.855,0.432,0.336,0.308,0.473          & Field \\
  V225    & 13.098056  &56.708333 &16.514$\pm$0.014 &  15.475$\pm$0.014 & -7.165  &  -2.929 &     0. &   0.053    &0.455,0.481,0.445                            & Field \\
  V226    & 13.025639  &56.589139 &16.803$\pm$0.020 &  15.733$\pm$0.014 &-12.308  &   3.219-&     0. &   0.060    &0.862,0.536                                  & Field \\
  V227    & 13.107472  &56.716111 &16.445$\pm$0.016 &  15.694$\pm$0.012 & -0.901  &    2.74 &     0. &   0.041    &0.528,0.494,0.474,0.486,0.476                & Field \\
  V228    & 13.044389  &56.622611 &17.311$\pm$0.032 &  16.127$\pm$0.012 & -5.825  &  -0.856 &   0.19 &   0.078    &0.477,0.858                                  & Field \\
\hline
\end{tabular}
\end{table*}
%***********************************************************************************************************************

\begin{figure*}
\hbox{
\includegraphics[width=9cm, height=9cm]{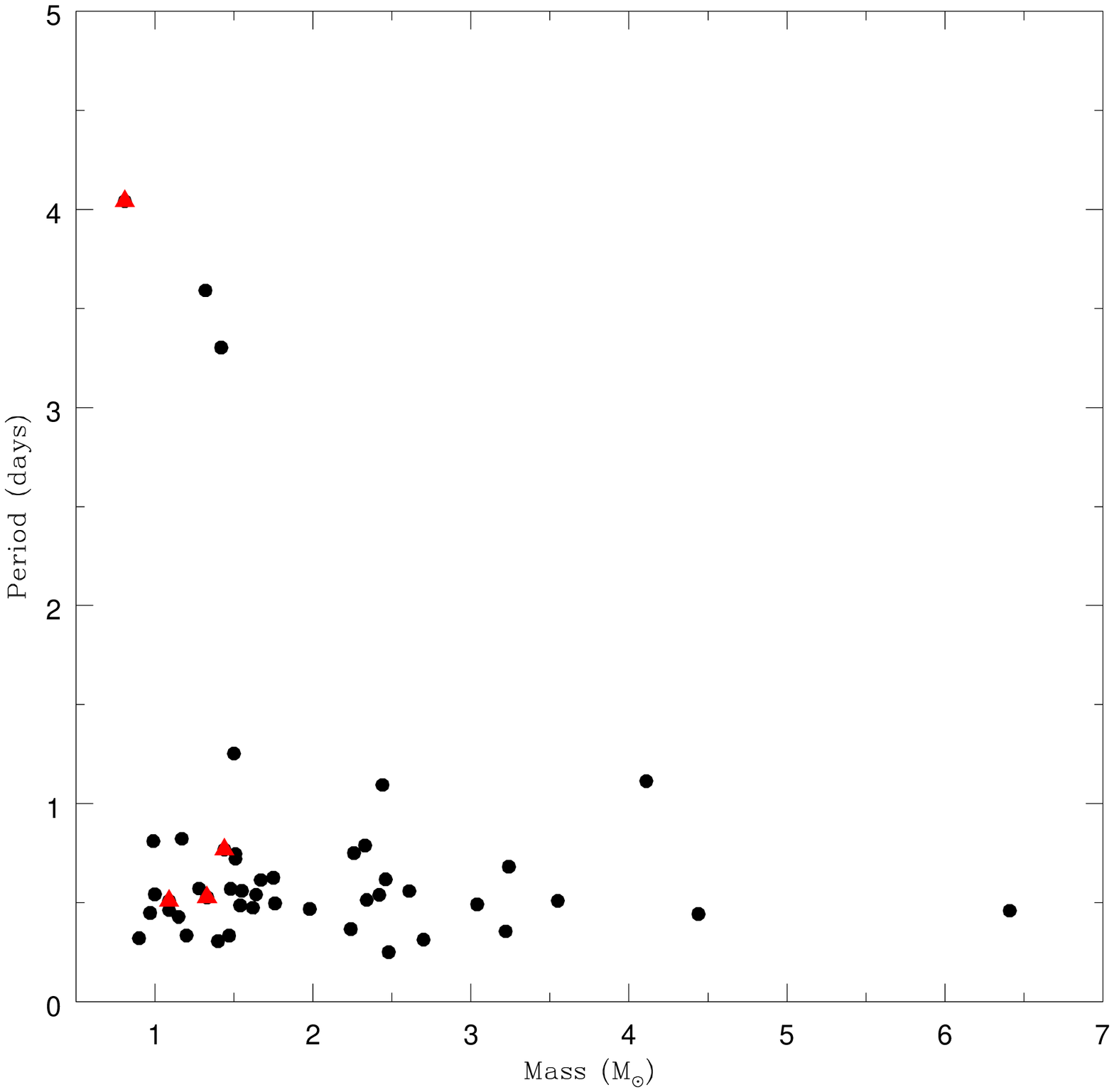}
\includegraphics[width=9cm, height=9cm]{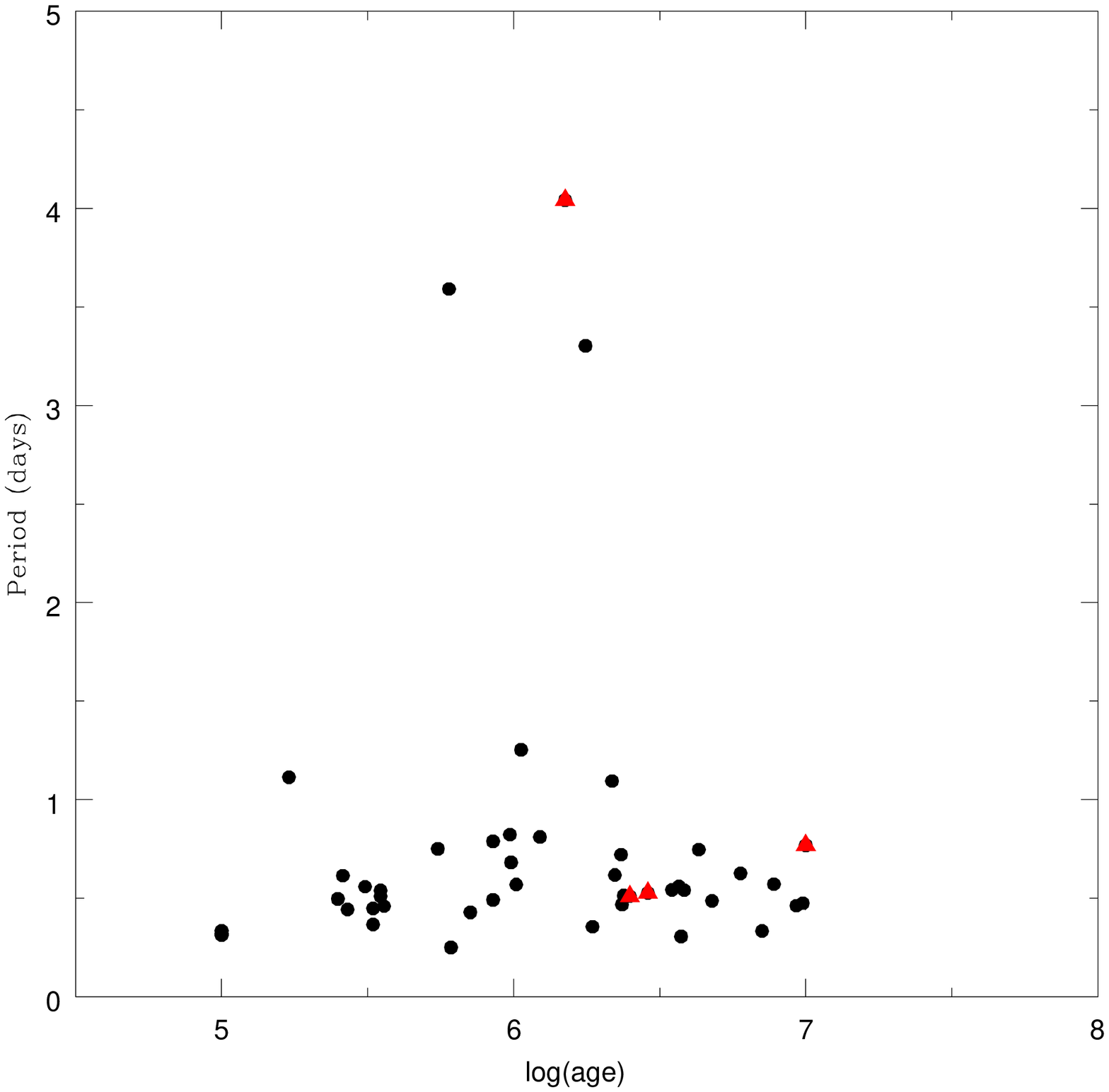}
}
\caption{Rotation period of TT variables as a function of mass and age. Filled 
circles represent WTTSs while triangle shows CTTSs.
}
\end{figure*}

\section{Nature of variable stars}
Of 228 periodic variables detected in the present work, 81 (30 MS and 51 PMS) stars are found to be 
associated with the cluster. The MS variable stars are classified according to their periods of variability, 
shape of light curves, as well as their locations in the Hertzsprung - Russell (HR) diagram. The $H-R$ diagram 
for MS variables is shown in Fig. 12. The location of variable stars in the  $H-R$ diagram
has been estimated with the help of their intrinsic colour
index $(B-V)_{0}$ and and $V$ magnitudes. The $U$, $B$, and $V$
magnitudes are available for 29 MS periodic variables. The $U$
data of V145 is not available. The intrinsic colour $(B-V)_{0}$
of variable stars has been estimated using the Q-method
(Guti$\acute{e}$rrez-Moreno 1975). For determination of luminosity
and temperature we require $V$ magnitudes, distance of the
cluster and relations provided by Torres (2010). The details
of the procedure for obtaining luminosity and temperature
of the star can be found in Lata et al. (2019). We have plotted
$(L/L_{\odot})$ versus temperature diagram for 29 MS variables in Fig. 12. Thus, based on location of 
variables in the $H-R$ diagram we have identified one
variable as $\beta$ Cep candidate. The $\beta$ Cep stars are massive
stars which are found to be located near the MS in the $H-R$
diagram. Their spectral types range from O to B. These variables pulsate with short periods. 
Multiple periods are often
found in these stars (Jager et al. 1982). Star V167 is detected as $\beta$ Cep candidate with a period about 0.2 days in the 
present work. Five slowly pulsating
B-type (SPB) candidates were also identified in the present work.
SPB stars as their name suggests, are B
spectral type stars. Their typical periods are in the range from 0.5 to 5 days
  and sometimes they show multiperiodicity.
Waelkens (1991) studied SPB stars and named these 
early-type variables as `SPB stars'. 
 The characteristics (periods and shape of light curves) of present variables detected as SPB stars are consistent with the class.     
The present sample of MS variables also consists of 15 new
class variables among MS variable members. The
new class variable stars have properties similar to the ones
discovered in NGC 3766 by Mowlavi et al. (2013).
 The periods and amplitudes of new class variable stars detected by Mowlavi et al. (2013) range from 0.1 to 0.7 days and 1 to 4 mmag, respectively.
The present new class variable stars have periods between 0.25 to 0.80 days.

Additionally, we have also detected 4 stars which could be
$\delta$ Scuti type variables. Their periods are found to be in the range of $\sim$ 0.22 - 0.25 days. These $\delta$ Scuti stars are found to be
pulsators which are located on/near the MS in the classical cepheid instability strip. 
These stars pulsate with period upto 5 or 6 hours. Three stars V94, V100 and V123 are
neither lying on any instability regions of $H-R$ diagram
nor in the location of new class variable. These stars might
be field stars or belong to the PMS population.

\begin{figure*}
\hbox{
\includegraphics[width=9cm, height=9cm]{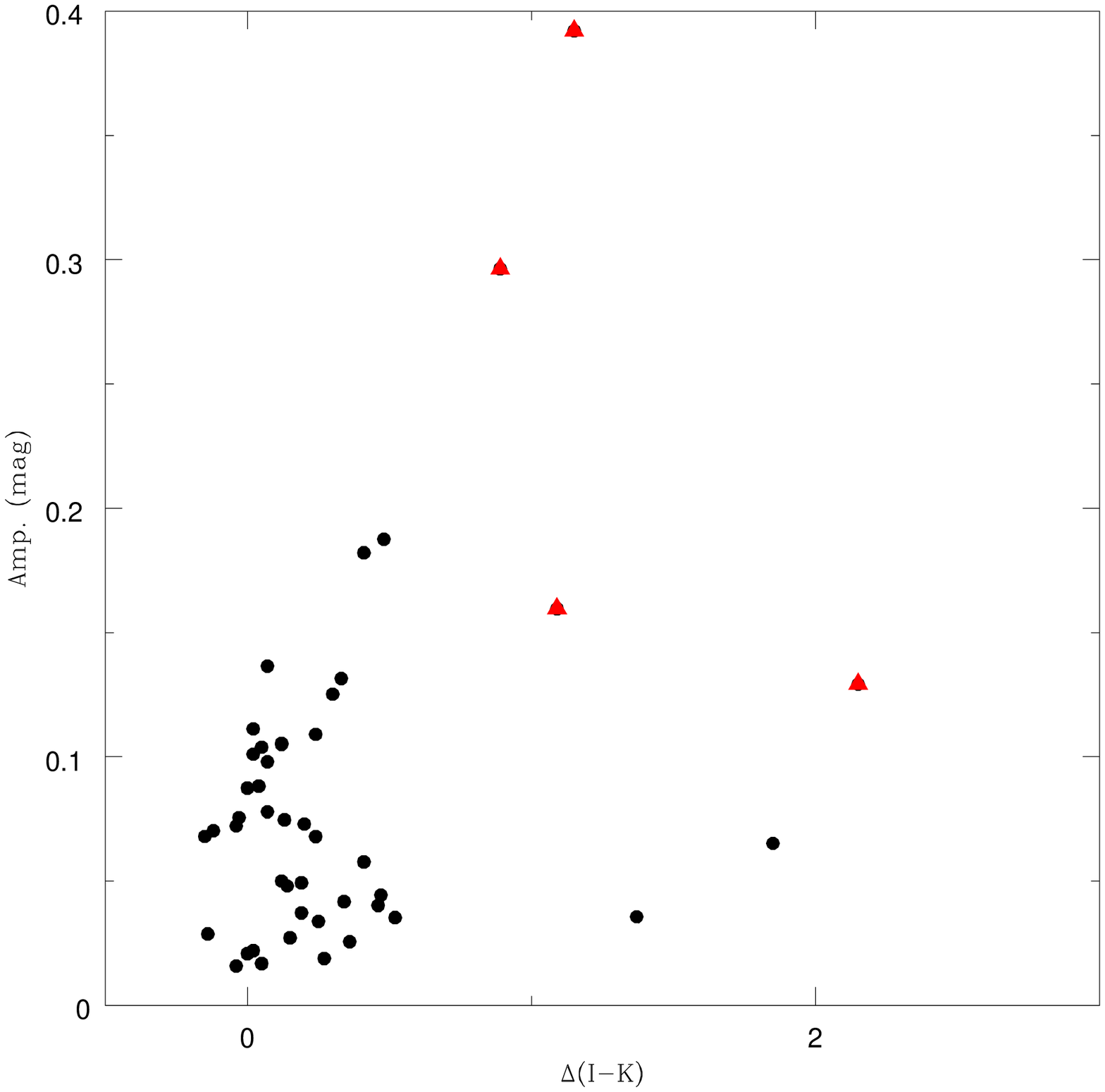}
\includegraphics[width=9cm, height=9cm]{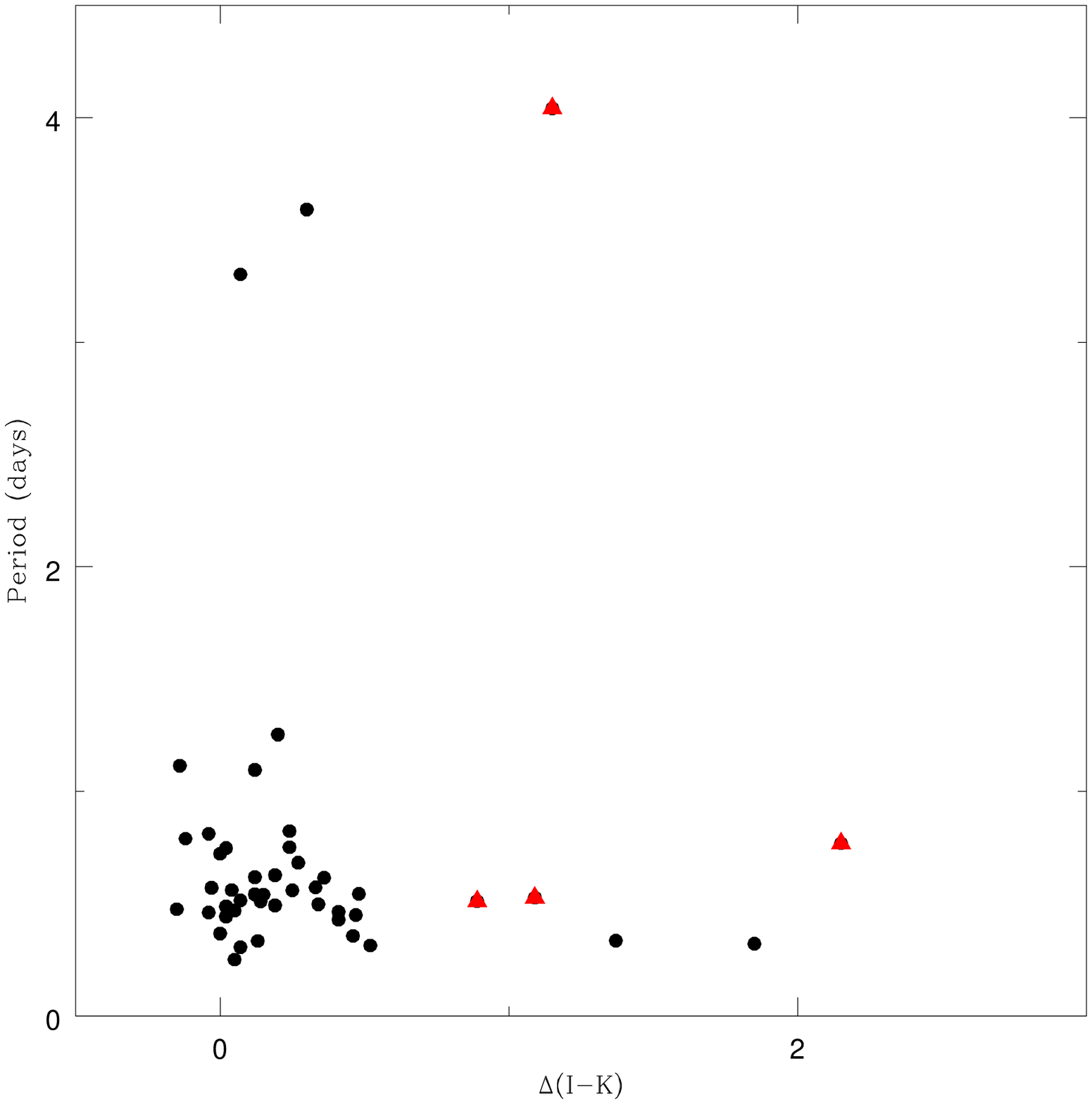}
}
\caption{Amplitude and rotation period of TT variables versus $\Delta (I-K)$. The symbols are same as in Fig. 13.
}
\end{figure*}

In the present study we have detected 51 PMS stars.  In general, the PMS objects are classified as T Tauri stars (TTSs) and Herbig Ae/Be stars. 
The TTSs are very young low mass stars  (mass $\lesssim$ 3$M_{\odot}$) which are still contracting and moving towards the MS (Herbig 1957, 1977), 
whereas  PMS stars of intermediate mass ($\gtrsim$3-10$M_{\odot}$) with emission lines are considered as Herbig  Ae/Be stars (Herbig 1960; Strom et al. 1972; Finkenzeller \& Mundt 1984). The TTSs are further classified  as Weak line TTSs (WTTSs) and 
Classical TTSs (CTTSs) on the basis of the strength of the $H{\alpha}$ emission line (Strom et al. 1989) which is measured by its equivalent width (EW). The WTTSs show weak  $H{\alpha}$ (EW$\le$10$\AA$) emission with almost negligible infrared excess, whereas CTTSs generally have a strong $H{\alpha}$ emission line with EW$>$10$\AA$, large ultraviolet and infrared excesses. 
 Both CTTSs and WTTSs show strong variability across all wavelengths and their brightness varies from a few minutes to years (see e.g., Appenzeller \& Mundt 1989). The brightness variation of TTSs are thought to occur  most probably due to the presence of cool and/or hot spots on their surface (Herbst et al. 1987, 1994), which may originate due to several mechanisms such as circumstellar disk material, accretion and magnetic field (Herbst et al. 1994). 

The cool spots on the surface of the stars are produced due to stellar magnetic fields on the photosphere and  rotate with the star. If these spots are symmetrically distributed over the photosphere, periodic brightness variations are observed in the light curves of the stars. The cool spots on the photosphere of stars are responsible for brightness variation in WTTSs and these objects are found to be fast rotators as they have either thin or no circumstellar disk. WTTSs are characterized by smaller stellar flux variations (a few times 0.1 mag).

Accreting CTTSs, surrounded by circumstellar  disks and  have hot spots on the surface produced by accreting material from disks on to the stars, show a complex behaviour in their optical and NIR light curves (Scholz et al. 2009). Irregular or non-periodic variations are produced because of changes in the accretion rate. The time-scales of brightness variation range from hours to years. Although the hot spots cover a smaller area on the stellar surface, their higher temperature produces larger variability amplitude (Carpenter, Hillenbrand \& Skrutskie 2001). The Herbig Ae/Be stars also show variability as they cross the instability strip in the HR diagram on their way to the MS. 
However, exact cause of variability in Herbig Ae/Be is still not known.

Of 51 probable PMS stars identified in the present work, 42, 4 and 5 are classified as WTTSs, CTTSs and Herbig Ae/Be stars, respectively. The amplitudes of WTTSs ranges from $\sim$0.015 mag to $\sim$0.19 mag, while periods of most WTTSs are found in the  range of $\sim$0.25 to $\sim$1.12 days. The periods and amplitudes of  CTTSs are found to vary $\sim$0.48 to $\sim$4 days and  $\sim$0.13 to $\sim$0.39 mag, respectively. The above  results suggest that stars with disk i.e., CTTSs have relatively larger amplitudes than the WTTSs and indicate that variability in CTTSs could be due to the presence of hot spot  on the stellar surface (see e.g., Carpenter, Hillenbrand \& Skrutskie 2001, Grankin et al. 2007, 2008, Lata et al. 2019). Similar results have been found in our earlier studies (Lata et al. 2019, 2016; Sinha et al. 2020).

Since the present sample of PMS stars is dominated by WTTSs (43 stars), we will further  focus our analysis to understand the evolution of rotation period and amplitude of WTTSs identified  in the present work. 
The masses and ages of PMS variables were estimated by comparing present observations with
the theoretical models to study the evolution of amplitude and period of PMS stars. The procedure for estimation of mass and age is given in Lata et al. (2019) and Chauhan et al. (2009).
For this, we have used PMS isochrones of Siess et al. (2000) in the age range of 0.1 to 10 Myr with an interval of 0.1 Myr and $V/V-I$ CMD of the cluster.
These theoretical isochrones corrected for the distance and reddening were
compared with the locations of PMS stars in the $V/V-I$ CMD.
Finally, we determine the mass and age of the PMS star corresponding to the closest isochrone on the CMD.
The dependence of rotation period of TTSs (i.e., WTTSs and CTTSs) on age and mass is shown in Fig. 13 which reveals that rotation period of WTTSs does not depend  either on age or mass of the stars.
However, the slowest rotators in the present study are low mass ($\sim$1$M_{\odot}$) stars.

WTTSs have either  thin or no circumstellar disk, it will be useful to estimate the influence of circumstellar disks on period and amplitude of these variable stars. The studies available in the literature e.g., Herbst et al. (2000); Littlefair et al. (2005); Cieza \& Baliber (2007); Cody et al. (2018), have used various disk indicators, such as EW of the H${\alpha}$ emission line and Ca II triplet lines, $\Delta(H-K)$ and $\Delta(I-K)$ excess, disk fraction, etc. In the present study we have used $\Delta(I-K)$ excess which is defined as below (cf. Hillenbrand et al. 1998);

$\Delta(I-K) = (I-K)_{obs}-(A_{I}-A _{K})- (I-K)_{0}$

where $(I-K)_{obs}$, $(I-K)_{0}$,  $A_{I}$  and $A_{K}$ are the observed colour, intrinsic colour of the star, interstellar extinction in the $I$ and $K$ bands, respectively.  The $A_{V}$ value is taken as 0.99 mag as determined using the relation $A_{V}/E(B-V)$. The values of $A_{I}$ and $A_{K}$ are estimated using the relations given by Cohen et al. (1981). To obtain intrinsic ($I-K)_{0}$ of variables we have used PMS evolutionary models of Siess et al. (2000) of a given mass and age. Fig. 14 displays the amplitude and period of TT variables as a function of $\Delta(I-K)$.  It is evident from Fig. 14 that the majority of WTTSs have $\Delta(I-K) < 0.3$ mag, hence they can be considered as disk-less sources. Apparently, no influence of $\Delta(I-K)$ either on amplitude or rotation of WTTSs is noticed.  
 It is worthwhile to mention that a correlation  between  $\Delta(I-K)$ and amplitude as well as period of CTTSs in the sense that larger values of the disk indicator, i.e.,  $\Delta(I-K)$ corresponds to relatively larger amplitude variations  has been reported in  previous studies (e.g., Dutta et al. 2018, Sinha et al. 2019).
 Similarly, larger values of $\Delta(I-K)$ indicates longer rotation periods.
An extensive study of ONC carried out by Herbst et al. (2000, 2002) reveals a strong correlation between rotation period and infrared excess, suggesting that the observed rotation period distribution could be due to the disk-locking mechanism.
Similar results were reported by Edwards et al. (1993) and the physical interpretation proposed by the authors was that the disks slow the rotation of stars through magnetic interaction (Koenigl 1991; Ostriker \& Shu 1995). An increasing trend in disk fraction with period in ONC and NGC 2264 was reported by Cieza \& Baliber (2007). However, Littlefair et al. (2005) did not find any correlation between the H${\alpha}$ equivalent width or $(K-L)$ excess and the rotation period of the stars.

To check the dependence of $\Delta(I-K)$ on the stellar mass, we plot these parameters for all the PMS variables identified  in the present study in Fig. 15 which reveals  that higher values of NIR excess are associated with relatively low mass stars. This is consistent with the result that the slowest rotators in the present study are low mass  ($\sim$1.0 $M_{\odot}$)  CTTSs  (cf. Fig. 13).

Fig. 16 plots amplitude of variability as a function of mass of the PMS variables, which manifests that
the amplitude decreases with the increase in mass. 

Fig. 17 displays amplitude as a function of age. It is interesting to note that the amplitude of variability of WTTSs show an increase for relatively older stars. An inspection of figure 20 by Sinha et al. (2019) also indicates that the youngest WTTSs (age $<$1 Myr) in the Sh 2-170 H II region also show lower value of amplitudes as compared to older WTTSs. 
In the inset of Fig. 17 we plot data for WTTSs from our previous studies for the clusters Be 59, NGC 1893, NGC 7380 
and Stock 8 (Lata et al. 2011, 2012, 2016 and 2019). The same trend has been noticed in previous data also.
Grankin (1999, 2013) and Grankin et al. (2008) have shown that a small periodicity amplitude suggests a more uniform distribution of spots over the stellar surface, while a large amplitude is typical of
the case where the spots are concentrated in one or two high-latitude regions, i.e., they are
distributed highly non uniformly. These conclusions
are  also confirmed by the Doppler mapping of the surfaces of selected PMS stars (Grankin 2013). Thus, Fig. 17 seems to suggest that configuration/distribution of spots on the photosphere of WTTSs changes as they become older.
The smaller amplitude for relatively larger mass ($\gtrsim$2.5$M_{\odot}$) as noticed in Fig. 16 could be either due to dispersal of disk
or uniform distribution of spots over stellar surface. It has been reported in our earlier studies that
the disk dispersal mechanism is less efficient for relatively low mass stars (Lata et al. 2011, 2012, 2016).

\begin{figure}
\includegraphics[width=9cm, height=9cm]{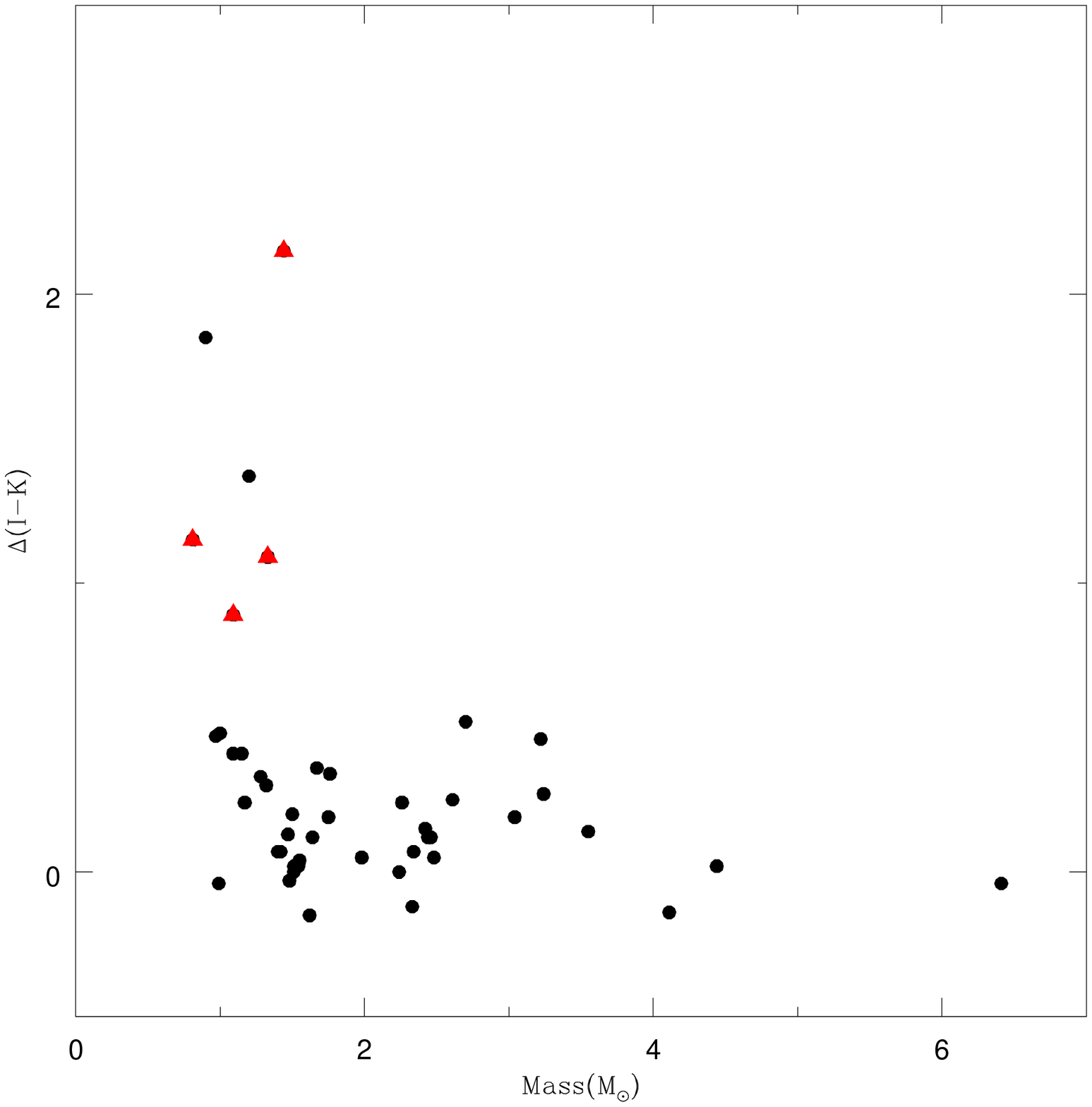}
\caption{Excess $\Delta (I-K)$ as a function of mass for TT variables.
The symbols are same as in Fig. 13.
}
\end{figure}

\begin{figure}
\includegraphics[width=9cm, height=9cm]{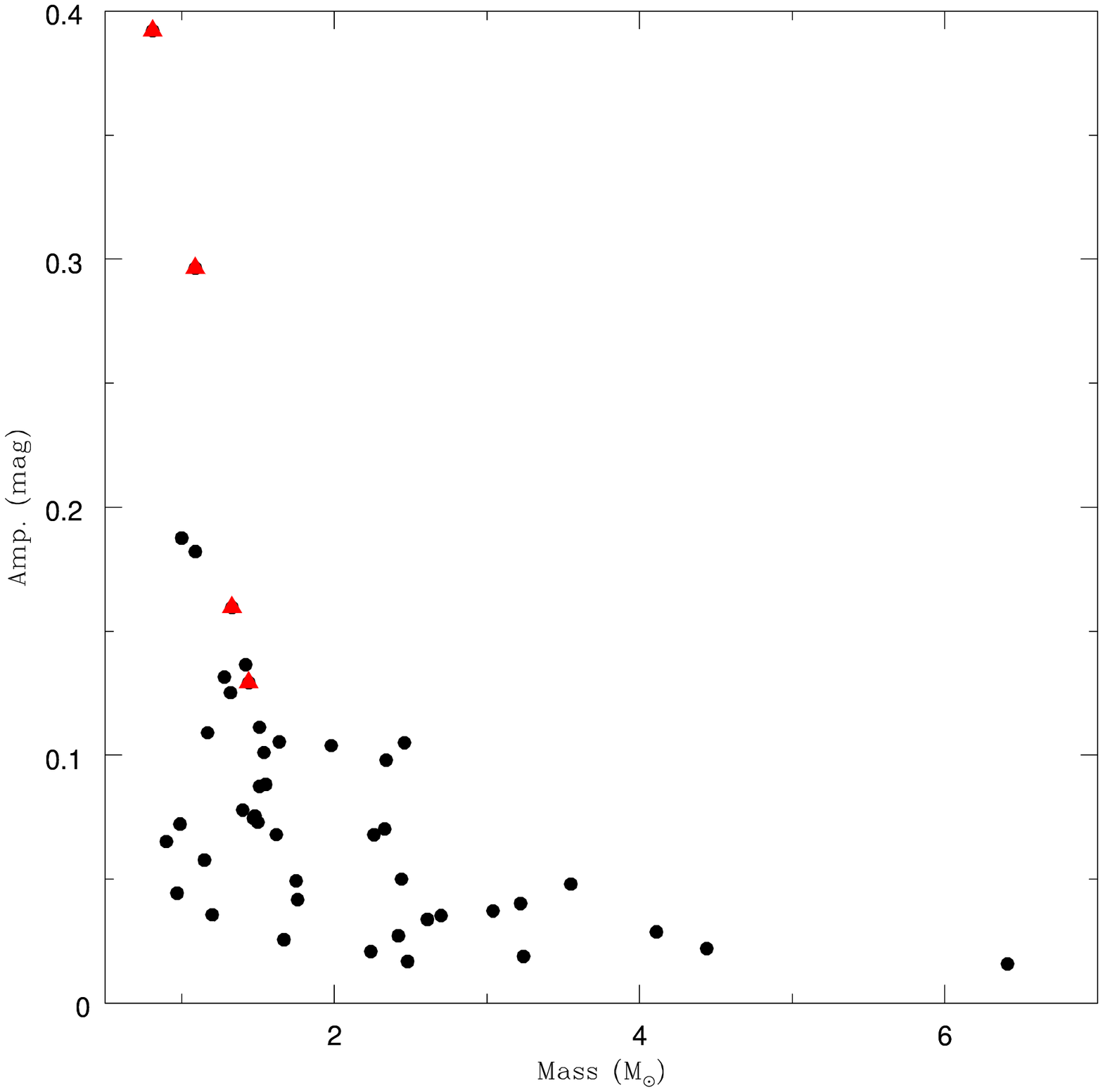}
\caption{Amplitude of present TT variable candidates as a function of mass.
The symbols are same as in Fig. 13.
}
\end{figure}

\begin{figure}
\includegraphics[width=9cm, height=9cm]{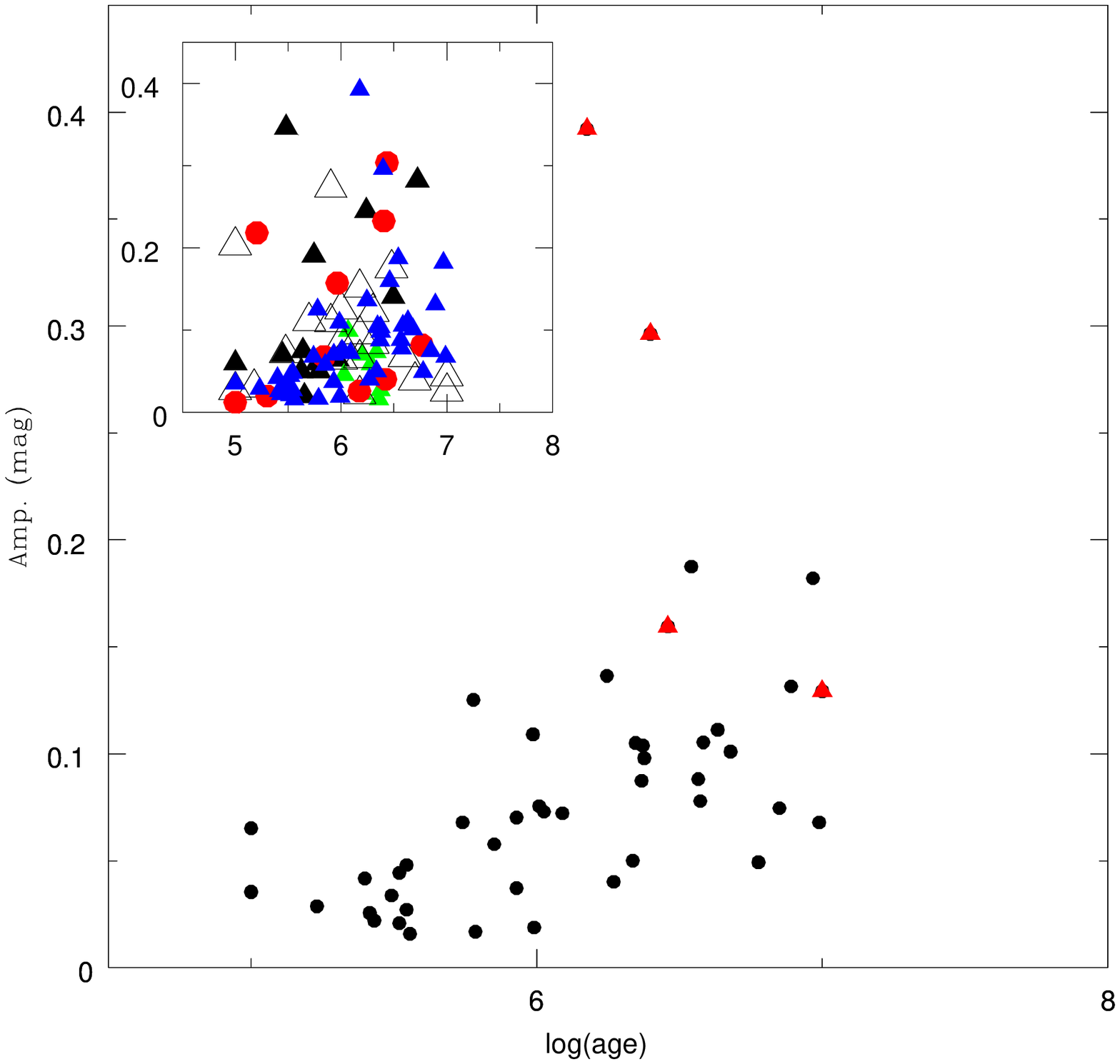}
\caption{Amplitude of present TT variable candidates as a function of age.
The symbols are same as in Fig. 13. Inset plots data for WTTSs taken from our earlier works (Lata et al. 2011, 2014, 2016 and 2019).
}
\end{figure}

The present study consists of 147 variables which could belong to the field population in the direction of NGC 281. 
Based on their light curves and variability characteristics, these stars could be RR Lyrae, $\delta$ Scuti, or binaries type variables.
As present study contains a few interesting variable stars which need to be characterized individually.
 
Star numbered V6 has a period of 0.157 days and an amplitude of 0.103 mag which indicates that the star might be a $\delta$ Scuti variable. 
The shape of the light curve also resembles the light curve of $\delta$ Scuti variables having nearly sinusoidal nature.

Star V15 has $V$ magnitude about 18.81 mag and it might belong to the field star population. With a period of 0.129 days and an amplitude of 0.261 mag, variable star V15 is suspected to be  of the Ellipsoidal variables type 
which is a subgroup of the rotating variable stars. The light curve is W-shaped with minima portion 
substantially broader than the maxima (peak) which is quite sharp. The two clearly visible minimas might not be equally deep due to stronger limb-darkening effect on the pointed portion of the more elongated star.

Star V42 has a period of 0.735 days and an amplitude of 0.066 mag. It could possibly be RR Lyrae (RRab) type variable since 
it has a period which is similar to RRab type variables. In addition to this, it has asymmetric shaped light curve which is 
again a characteristic of RR Lyrae type variables.

Star V123 has a period of 0.76 days and amplitude of 0.049 mag. This star may be Cataclysmic variable of nova type as the light curve indicates that the amplitude decreases gradually as in case of cataclysmic variable.

Star V126 has a period of 0.06 days and 0.531 days with an amplitude of 0.015 mag. It is considered as PMS star (Herbig Ae/Be) in the present work. 
The shape of the light curve of this star indicates that 
this might be RS Canum Venaticorum eclipsing binary. In addition to this, amplitude of such variables are in the range of 0.01 to 0.6 mag 
and the amplitude of this star also comes in this range. 

In the present work star V152 is considered as a PMS star.
Sharma et al. (2012) identified star V152 as a PMS source. However, this star in $J-H/H-K$ TCD does not lie in the region of class I, class II, II or Herbig Ae/Be.
In fact, this is a bright star of $V$ magnitude around 11.3 mag, and in $V/V-I$ CMD it is located on/near MS with other identified MS stars.
It has a period of 0.46 days and amplitude of 0.016 mag, and  
its variability characteristics also reveals that it could be a slowly pulsating star. 

The period of star V153 and V168 is 0.111 and 0.105 days, respectively and their amplitude is found as 0.018 mag. Based on variability characteristics, 
these stars could be of $\delta$ Scuti type variables. 

\section {Summary}
The present work detected 228 periodic variables in the field where young open cluster NGC 281 is located. The association of variables to 
the cluster has been discussed on the basis of TCDs, CMD and kinematic data. The 
membership probabilities of 223 stars have been determined using Gaia proper motion data.  
Using the present time series photometric data we have detected a number of PMS, MS and field variable stars.
Eighty one of 228 variables are members, with 30 MS members and 51 PMS members. Based on their periods and light curves shape, and locations in the $H-R$ diagram we have characterized MS 
variable stars into different types of variability. 
These MS variables are classified as $\beta$ Cep, $\delta$ Scuti, SPB and new class type variables. 
The identified PMS variable stars are characterized as CTTSs, WTTSs and Herbig Ae/Be stars.
The present study also indicates that the CTTSs vary with larger amplitude in comparison to the WTTSs.
The masses and ages of PMS stars have been derived using CMD and PMS theoretical models.
It is found that rotation period of WTTSs does not depend either on age or mass of the stars.
Amplitude and mass of TT variables are found to be correlated in the sense that relatively massive stars ($\gtrsim$2.5$M_{\odot}$) 
have smaller amplitudes. This could be due to lack of disk or uniform distribution of spots on the photosphere of the stars. 
 We also note that the amplitude of variability of WTTSs increases with age, suggesting that the configuration/distribution of spots on the surface of WTTSs changes with their age.
 There are 147 variables which could belong to the field population. These field variables could be RR Lyrae, $\delta$ Scuti, or binaries type variables.

\section{Acknowledgment}
We are thankful to the referee for careful reading of the paper and valuable suggestions/comments.

\section*{Availability of data}
The data underlying this article will be shared on reasonable request to the corresponding author.

The data underlying this article are available in 2MASS (https://vizier.u-strasbg.fr/viz-bin/VizieR?-source=II/246) and Gaia (https://gea.esac.esa.int/archive/).

\bibliographystyle{mn2e}

\begin{thebibliography}{36}
\expandafter\ifx\csname natexlab\endcsname\relax\def\natexlab#1{#1}\fi
\bibitem[]{}Appenzeller I., \& Mundt R., 1989,  Astron. Astrophys. Rev.  1, 291
\bibitem[]{}Arcos C., Kanaan S., Chavez J., Vanzi L., Araya I., Cure M.,  2018, MNRAS, 474, 5287
\bibitem[]{}Balona L. A., Pigulski A., Cat P. De, Handler G., Guti$\acute{e}$rrez-Soto J., Engelbrecht C. A., Frescura F., Briquet M., et al., 2011, MNRAS, 413, 2403
\bibitem[]{}Bessell M. S., Brett J. M., 1988, PASP, 100, 1134
\bibitem[]{}Burger M., de Jager C., van den Oord, G. H. J., Sato N., 1982, A\&A, 107, 320
\bibitem[]{}Carpenter J. M.,  Hillenbrand L. A.,  Skrutskie M. F., 2001  AJ, 121, 3160
\bibitem[]{}Cabrera-Cano J. and Alfaro E. J., 1990, A\&A, 235, 94
\bibitem[]{}Chauhan N., Pandey A. K., Ogura K. et al, 2009, MNRAS, 396, 964
\bibitem[]{}Cutri R. M.  et al., 2003 , 2MASS All Sky Catalog of Point Sources, VizieR Online Data Catalog, University of Massachusetts and Infrared Processing and Analysis Center (IPAC/California Institute of Technology), 2246, 0 https://vizier.u-strasbg.fr/viz-bin/VizieR?-source=II/246
\bibitem[]{}Cohen J. G., Persson S. E., Elias J. H., Frogel J. A., 1981, ApJ, 249, 481
\bibitem[]{}Cody A. M., Hillenbrand L. A. 2018, AJ, 156, 71
\bibitem[]{}Cieza Lucas, Baliber Nairn, 2007, ApJ, 671, 605
\bibitem[]{}Dutta S., Mondal S., Joshi S., Das R., 2019, BSRSL, 88, 103
\bibitem[]{}de Jager C., Sato, N., Burger M., Neven L., 1982, Ap\&SS, 83, 411
\bibitem[]{}Edwards S., Strom S. E., Hartigan P., Strom K. M., Hillenbrand L. A., Herbst W., Attridge J., Merrill K. M., Probst R., Gatley I., 1993, AJ, 106, 372
\bibitem[]{}Elmegreen B. G. \& Lada C. J.,  1978, ApJ, 219, 467
\bibitem[]{}Finkenzeller U., Mundt R., 1984, A\&AS, 55, 109
\bibitem[]{}Gaia Collaboration et al. 2018, A\&A, 616, 13 https://gea.esac.esa.int/archive/
\bibitem[]{}Getman K. V., Feigelson E. D., Sicilia-Aguilar A., et al., 2012, MNRAS, 426, 2917
\bibitem[]{}Girardi L., Bertelli G., Bressan A., Chiosi C., Groenewegen M. A. T., Marigo P., Salasnich B., Weiss A., 2002, A\&A, 391, 195
\bibitem[]{}Grankin K. N.,  Melnikov S. Yu.,  Bouvier J.,  Herbst W., Shevchenko V. S., 2007, A\&A, 461, 183
\bibitem[]{}Grankin K. N.,  Bouvier J.,  Herbst W.,  Melnikov S. Y., 2008, A\&A, 479, 827
\bibitem[]{}Grankin K. N.,  2013, Astronomy Letters, 2013, Vol. 39, 446
\bibitem[]{}Grankin K. N., 1999,  Astron. Lett. 25, 526 
\bibitem[]{}Guti$\acute{e}$rrez-Moreno A.,  1975, PASP 87, 805
\bibitem[]{}Herbig G. H., 1960, ApJS, 4, 337
\bibitem[]{}Herbst W.,  Herbst D. K.,  Grossman E. J.,  Weinstein D., 1994, AJ, 108, 1906
\bibitem[]{}Herbst W., Booth J. F., Korett F. L., et al, 1987, AJ, 94, 13
\bibitem[]{}Herbst W., Bailer-Jones C. A. L., Mundt R., Meisenheimer K., Wackermann R., 2002, A\&A, 396, 513
\bibitem[]{}Herbst W., Maley J. A., Williams E. C., 2000, AJ, 120, 349
\bibitem[]{}Ivers Carol B., Booker M., Piper M., Powers L., Ali B., Wolk S. J., 2014, NITARP, AAS, 22324419
\bibitem[]{}Jose J., Pandey A. K., Samal M. R., Ojha D. K., Ogura K., Kim J. S., Kobayashi N., Goyal A., Chauhan N., Eswaraiah C., 2013, MNRAS, 432, 3445
\bibitem[]{}Koenigl Arieh, 1991, ApJ, 370L, 39 
\bibitem[]{}Landolt A. U., 1992, AJ, 104, 340
\bibitem[]{}Lata S., Pandey A. K., Maheswar G., Mondal S. and Kumar B., 2011, MNRAS, 418, 1346
\bibitem[]{}Lata S., Pandey A. K., Chen W. P., Maheswar G. and Chauhan N., 2012, MNRAS, 427, 1449
\bibitem[]{}Lata S., Pandey A. K., Panwar N. et al., 2016, MNRAS, 456, 2505
\bibitem[]{}Lata Sneh, Pandey Anil K., Kesh Yadav Ram, et al. 2019, AJ, 158, 68
\bibitem[]{}Lomb N. R., 1976, Ap\&SS, 39, 447
\bibitem[]{}Littlefair S. P., Dhillon V. S., Mart$\acute{i}$n E. L., 2005, A\&A, 437, 637
\bibitem[]{}Meyer M. R., Calvet N., Hillenbrand L. A., 1997, AJ, 114 , 288
\bibitem[]{}Mowlavi N.,  Barblan F.,  Saesen S.,  Eyer L., 2013,  A\&A, 554, 108
\bibitem[]{}Ostriker Eve C., Shu Frank H., 1995, ApJ, 447, 813
\bibitem[]{}Sato M.  et al.,  2008, PASJ, 60, 975
\bibitem[]{}Scholz A., Eisl{\H o}ffel J., Mundt R., 2009, MNRAS, 400, 1548
\bibitem[]{}Scargle J. D., 1982, ApJ, 263, 835
\bibitem[]{}Sesar B. et al, 2007, AJ, 134, 2236
\bibitem[]{}Sharma Saurabh, Pandey Anil K., Pandey Jeewan C., Chauhan Neelam, Ogura Katsuo, Ojha Devandra K., Borrissova Jura, Mito Hiroyuki, Verdugo Thomas, Bhatt Bhuwan C., 2012, PASJ, 64, 107
\bibitem[]{}Siess L., Dufour E., Forestini M., 2000, A\&A, 358, 593
\bibitem[]{}Sinha Tirthendu, Sharma Saurabh, Pandey A. K., Yadav R. K., Ogura K., Matsunaga N., Kobayashi N., Bisht P. S., Pandey R., Ghosh A., 2020, MNRAS, 493, 267S 
\bibitem[]{}Stetson P. B.,  1987, PASP, 99, 191
\bibitem[]{}Stetson P. B.,  1992, J. R. Astron. Soc. Can., 86, 71
\bibitem[]{}Strom S. E., Strom K. M., Yost J., et al, 1972, ApJ, 173, L65
\bibitem[]{}Strom K. M., Strom S. E., Edwards S., Cabrit S., Skrutskie, M. F. 1989, AJ, 97, 1451
\bibitem[]{}Torres G., 2010,  AJ, 140, 1158
\bibitem[]{}Waelkens C., 1991, A\&A, 246, 453
\end{thebibliography}

\end{document}